%% file: 0.main.tex
\documentclass[sigconf]{acmart}
\usepackage{lipsum} 
\usepackage{graphicx}
\usepackage{colortbl}
\usepackage[colorinlistoftodos]{todonotes}
\usepackage{amsmath,mathtools}

\usepackage{array}
\usepackage{makecell}
\usepackage{threeparttable}
\usepackage[ruled,vlined,linesnumbered]{algorithm2e}
\usepackage{setspace}
\usepackage[skip=5pt]{caption}
\usepackage{subfig}
\usepackage{enumitem}
\usepackage{algorithmicx}
\usepackage{color}
\usepackage{titlesec}
\usepackage{pifont}
\usepackage{xcolor}
\usepackage{array}
\usepackage{svg}
\usepackage{amsmath}

\definecolor{mygreen}{RGB}{0,100,0}
\newcommand\acsac[1]{{\color{black} {#1}}}

\graphicspath{{Figures/}}
\DeclareGraphicsExtensions{.pdf,.png,.jpg,.eps} 

\algnewcommand{\IfThenElse}[3]{
  \State \algorithmicif\ #1\ \algorithmicthen\ #2\ \algorithmicelse\ #3}

\setcopyright{none} 
\acmYear{2022}

\begin{document}
\title{HWGN\textsuperscript{2}: Side-channel Protected Neural Networks through Secure and Private Function Evaluation}

\author{Mohammad Hashemi}
\email{mhashemi@wpi.edu}
\affiliation{\institution{Worcester Polytechnic Institute}
\city{Worcester, MA}
\country{USA}}
\author{Steffi Roy}
\email{steffiroy@ufl.edu}
\affiliation{\institution{University of Florida}
\city{Gainesville, FL}
\country{USA}}
\author{Domenic Forte}
\email{dforte@ece.ufl.edu}
\affiliation{\institution{University of Florida}
\city{Gainesville, FL}
\country{USA}} 
\author{Fatemeh Ganji}
\email{fganji@wpi.edu}
\affiliation{\institution{Worcester Polytechnic Institute}
\city{Worcester, MA}
\country{USA}}


\begin{abstract}
Recent work has highlighted the risks of intellectual property (IP) piracy of deep learning (DL) models from the side-channel leakage of DL hardware accelerators.
In response, to provide side-channel leakage resiliency to DL hardware accelerators, several approaches have been proposed, mainly borrowed from the methodologies devised for cryptographic implementations. 
Therefore, as expected, the same challenges posed by the complex design of such countermeasures should be dealt with. 
This is despite the fact that fundamental cryptographic approaches, specifically secure and private function evaluation, could potentially improve the robustness against side-channel leakage. 
To examine this and weigh the costs and benefits, we introduce hardware garbled NN (HWGN\textsuperscript{2}), a DL hardware accelerator implemented on FPGA. 
HWGN\textsuperscript{2} also provides NN designers with the flexibility to protect their IP in real-time applications, where hardware resources are heavily constrained, through a hardware-communication cost trade-off. 
Concretely, we apply garbled circuits, implemented using a MIPS architecture that achieves up to $62.5\times$ fewer logical and $66\times$ less memory utilization than the state-of-the-art approaches at the price of communication overhead. 
Further, the side-channel resiliency of HWGN\textsuperscript{2} is demonstrated by employing the test vector leakage assessment (TVLA) test against both power and electromagnetic side-channels. 
This is in addition to the inherent feature of HWGN\textsuperscript{2}: it ensures the privacy of users' input, including the architecture of NNs. 
We also demonstrate a natural extension to the malicious security model- just as a by-product of our implementation. \vspace{-5pt}
\end{abstract}




\maketitle

\input{1.Introduction}
\input{2.Background}

\input{3.Theory}

\input{3_1.Implementation}
\input{4.Practical_Implementation}
\input{6.Discussion}
\input{7.Conclusion}

\bibliographystyle{ACM-Reference-Format}
\bibliography{references}
\input{8.appendix}

\end{document}

%% file: 1.Introduction.tex
\section{Introduction}\label{sec:intro}
An ever-increasing number of applications are demanded from machine learning and, in particular, deep learning (DL). 
These applications include image processing, natural language processing, network intrusion detection systems (NIDSs), biometrics systems, and analysis of medical data -- just to name a few cf.~\cite{lecun2015deep}. 
These, among other compute-intensive services, have been supported by cloud platforms equipped with hardware acceleration~\cite{chakraborti2022cloud}; however, cloud platforms are not the only hosts of DL algorithms and modules. 
IoT edge devices have embodied modules to perform many tasks, for instance, image classification or speech recognition as required by wearable devices for augmented reality and virtual reality~\cite{lecun20191}.  
In addition to those, so-called mobile and wearable devices, low-cost DL chips (e.g., sensors or actuators) have been employed in cameras, medical devices, appliances, autonomous surveillance, ground maintenance systems, and even toys.
In these cases, heavy demands have been put on (always-on) \emph{DL-inference} accelerators with power constraints, e.g., ARM Cortex-M microcontrollers. 

Under DL-inference scenarios, trained neural networks (NNs) are made available to users. 
To obtain such a trained NN, a large training dataset is used in a time-consuming process to tune NN hyperparameters, which cannot be repeated in a straightforward manner. 
Therefore, it can be tempting for an adversary to target the DL-inference accelerator and extract those parameters. 
Besides hyperparameters, the architecture of NNs is another asset to protect as it may (even partially) reveal private information~\cite{gilad2016cryptonets,fredrikson2014privacy} or at least help the adversary to reconstruct the NN~\cite{ateniese2015hacking} cf.~\cite{batina2019csi}. 
\acsac{In fact, having the knowledge of the NN architecture allows the adversary to launch power attacks, for instance, cryptanalysis of trained NNs aiming to extract an identical copy of NN models~\cite{carlini2020cryptanalytic}. }
Since physical access can make it further easier for attackers to reverse-engineer and disclose the assets (i.e., architecture and hyperparameters) corresponding to NNs, usual protections, e.g., blocking binary readback, blocking JTAG access, code obfuscation, etc. could be applied to prevent binary analysis~\cite{batina2019csi}. 
These, of course, would not stop an attacker from leveraging the information that leaks through side-channels. 

Side-channel analysis (SCA) has been widely performed to evaluate the robustness of cryptographic implementations against one of the most dangerous types of physical attacks. 
SCA has been considered effective and applicable due to its (relatively) low cost and feasibility in real-world conditions, especially, when
applied against small embedded devices~\cite{standaert2010leakage}. 
Interestingly enough, DL has played an important role in the SCA domain of study by offering competitive performance corresponding to more capable attackers~\cite{picek2021sok}. 
In doing so, DL can be a double-edged sword: DL implementations have been already come under side-channel attacks~\cite{batina2019csi,xiang2020open,yu2020deepem,dubey2020maskednet}. 
These attacks have resulted in considerable efforts to devise countermeasures. 
Intuitively, masking schemes developed to protect cryptographic modules against SCA have been one of the first solutions discussed in the literature~\cite{dubey2020bomanet,dubey2022modulonet}. 
These methods come with their own set of challenges, e.g., being limited to a pre-defined level of security associated with the masking order or even to a particular modality. 
Moreover, evidently, masking cannot stop the attacker from disclosing the architecture of the NN under attack. 

The natural question to be asked is why fundamental cryptographic concepts that can provide NNs with robustness against SCA have not yet been examined. 
Concretely, secure function evaluation (SFE), specifically garbled circuits evaluation, has been considered to prevent side-channel leakage cf.~\cite{jarvinen2010garbled,mantel2020ricasi}. 
\acsac{Nevertheless, in practice, SFE has not been considered to stop side-channel attacks, perhaps, due to the high overhead initially observed in~\cite{jarvinen2010garbled}. 
This study has attempted to introduce a provably side-channel resistant one-time program~\cite{goldwasser2008one} implemented on a field-programmable gate array (FPGA). 
Their implementation is a combination of tamper-resistant hardware with Yao's garbling scheme~\cite{yao1986generate}, which comes with an overhead 
of about factor $10^6\times$ compared to an unprotected AES embedded in an FPGA. }

Apart from the leakage properties of SFE and its realization garbled circuits, they have been developed to ensure the security of users' data, when two parties jointly evaluate a known function. 
Therefore, in a natural way, garbled circuits have been investigated to put forward the notion of privacy-preserving inference-as-a-service~\cite{riazi2019xonn,riazi2018chameleon}. 
These results are further backed by more theoretical studies that aim to, for instance, heavily optimize garbled circuits to suit TensorFlow classifier descriptions~\cite{ball2019garbled}. 
In spite of these results, the gap between these studies is evident: design of countermeasures against SCA, software implementation of garbled NNs~\cite{riazi2019xonn,rouhani2018redcrypt,hussain2020tinygarble2}, and hardware implementations of garbled circuits~\cite{songhori2015tinygarble}. 
To narrow this gap, \acsac{this paper introduces HWGN\textsuperscript{2} (hardware garbled NN) }and contributes to the following aims. 
\vspace{-5pt}
\begin{itemize}[wide = 0pt]
    \item \textit{A secure and private DL-inference hardware accelerator\textcolor{black}{, resilient to SCA.}}
    \acsac{To protect the NN model (including its architecture and parameters) against SCA, HWGN\textsuperscript{2} relies on the principles of private function evaluation (PFE) and SFE, realized through a general purpose processor cf.~\cite{songhori2015tinygarble,songhori2016garbledcpu}. }
    \textcolor{black}{Interestingly enough, as opposed to the argument in~\cite{jarvinen2010garbled,mantel2020ricasi} suggesting the side-channel resiliency of garbled circuits, Levi et al. have recently demonstrated a side-channel attack against garbling schemes leveraging the free-XOR optimization~\cite{levi2022garbled}. 
    HWGN\textsuperscript{2} is not susceptible to this attack since PFE is taken into account to make the function private. }
    It is noteworthy that the privacy of the NN model is understudied even in existing software garbled  DL-inference~\cite{riazi2019xonn,rouhani2018redcrypt,ball2019garbled}. 
    Interestingly enough, our instruction set-based HWGN\textsuperscript{2} is model-agnostic. 
    Moreover, in the most cost-efficient setting with a DL-inference realized by using XNOR operators, our implementation does not require any modification to the NN, in contrast to what has been proposed as software garbled DL-inference~\cite{riazi2019xonn}. 
    \item \textit{Assessment of the effectiveness and cost of SCA protection relying on SFE/PFE.} 
    \acsac{To evaluate the feasibility of our approach, we identify two implementation scenarios, namely (1) resource- and (2) communication-efficient. 
    In the first category, compared to the unprotected NN, the overhead is up to $0.0011\times$ and $0.018\times$ more logical and memory hardware resources, respectively; however, this relatively low overhead is achieved at the cost  of communication between the user and the inference service provider.
    If communication constitutes a burden on the system, it can be dealt with, even though compared to the unprotected design, the overhead increases to $52.4\times$ and $40.8\times$ more logical and memory hardware resources, respectively.
    However, even under the communication-efficient scenario, HWGN\textsuperscript{2} utilizes up to $62.5\times$ fewer logical and $66\times$ less memory, respectively, compared to the most relevant study~\cite{rouhani2018redcrypt}. 
    Additionally, the side-channel resiliency of HWGN\textsuperscript{2} implementation on the FPGA is assessed by applying T-test leakage detection.  }
    \item \textit{Comprehensive discussion on various aspect of our design.} Although SCA better fits the specification of an attacker in the honest-but-curious (HbC) setting as defined in cryptography, it is also secure in the presence of malicious (active) adversary~\cite{lindell2016fast} as a byproduct of our hardware implementation. 
    Furthermore, we discuss how the implementation of maliciously-secure GCs can form a basis for allowing the user to perform multiple executions~\cite{huang2014amortizing}. 
\end{itemize}
\begin{figure}[t]
\centering \noindent
\includegraphics[width=0.95\columnwidth]{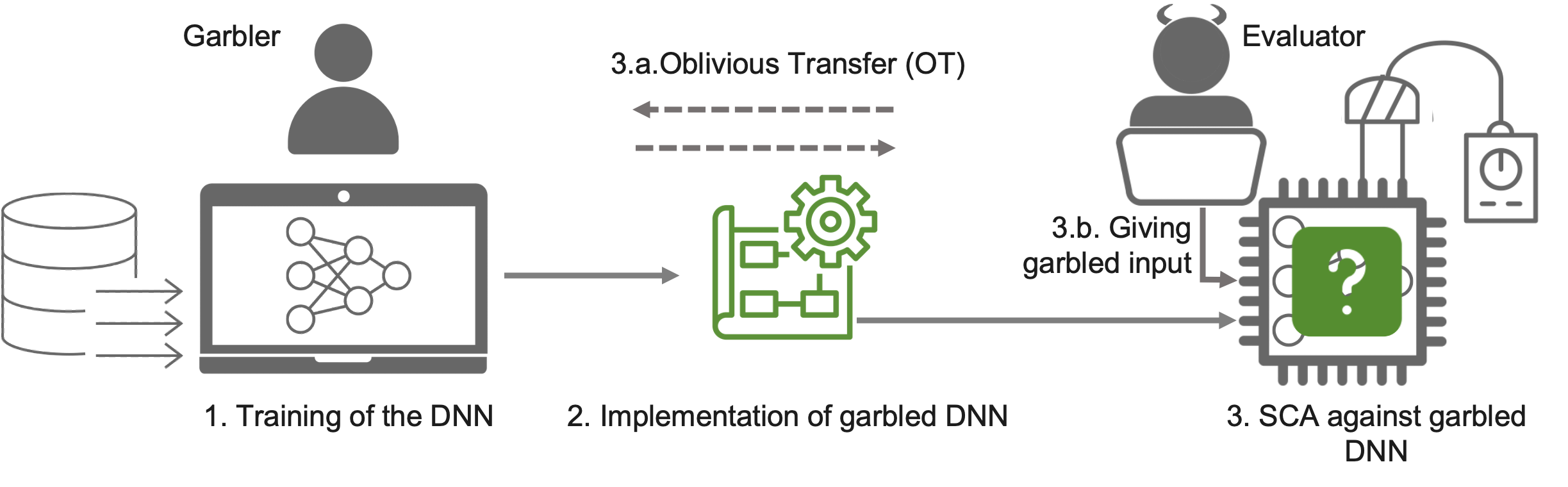}
\caption{HWGN\textsuperscript{2} framework: The process begins with training the NN as done for a typical DL task. 
The second step corresponds to the implementation of the garbled NN hardware accelerator along with running the OT protocol. 
The end-user poses the accelerator and attempts to collect the side-channel traces with the aim of extracting information on the NN (architecture, hyperparameters, etc.).}
\label{fig:framework}
\end{figure}

%% file: 2.Background.tex
\section{Adversary Model}\label{Adversary Model}
Valuable assets of NNs, as intellectual property (IP), include their NN architectures, hyperparameters, and the parameters critical to achieving reasonable accuracy~\cite{batina2019csi}. 
On the other hand, these NNs might be used in applications in which their inputs contain sensitive information (e.g., medical or defense records~\cite{mittal2021survey}).
Hence, the security of inputs given to NNs along with the privacy of the networks themselves must be guaranteed. 
Note that here the definitions of security and privacy are borrowed from SFE- and private function evaluation- (PFE) related literature~\cite{bellare2012foundations}. 
Classically, two threat models have been considered in prior works in the contexts of SFE and PFE: (i) \textit{semi-honest} (so-called Honest-but-curious (HbC)) and (ii) \textit{malicious} (active) adversary.
An HbC adversary is expected to follow the protocol execution and does not deviate from the protocol specifications.
To be more specific, the HbC adversary may only be able to learn information without interfering with the protocol execution.
On the contrary, a malicious adversary may attempt to cheat or deviate from the protocol execution specifications. 
In doing so, to extract the information on users' data, the adversary constructs an arbitrary, well-chosen garbled circuit computing a different function, which cannot be detected due to the garbling~\cite{lindell2016fast}.  
It is evident that this scenario does not match the one describing SCA; however, we will discuss how HWGN\textsuperscript{2} naturally ensures the security even in this setting, where NN developers would attempt to violate users' data security (see Section~\ref{sec:discussion}).  

In our work, we consider (mainly) the HbC adversary, whose role is played by Bob (i.e., the evaluator), whereas Alice is the garbler~\cite{lindell2016fast,bellare2012foundations} (see Figure~\ref{fig:framework}). 
Following the definition of the attack model presented in the state-of-the-art, e.g.,~\cite{dubey2022modulonet,dubey2020bomanet}, the garbler trains the DL model in an offline fashion, and the evaluator performs the inference. 
It is important to stress that the hardware implementation encompasses solely the evaluator engine, i.e., neither garbling module nor encryption one is implemented on the hardware platform. 
To evaluate the garbled DL accelerator, the evaluator feeds her garbled inputs prepared in an offline manner.    
The evaluator can collect power/EM traces from the device either via direct access or remotely, see e.g.,~\cite{schellenberg2018inside,zhao2018fpga}. 
For this, the evaluator follows a chosen-plaintext-type attack model, where she sends her inputs to the device for classification and readily captures multiple traces. 
These traces will be then used to launch power/EM-based side-channel attacks~\cite{kocher1999differential,chari2002template,brier2004correlation}. 
The goal of the garbler is to protect the NN architectures, hyperparameters, and parameters from the HbC evaluator. 
HWGN\textsuperscript{2} fulfills this requirement through SFE/PFE techniques (see Section~\ref{sec:foundation}). 
Moreover, in contrast to prior countermeasures against SCA mounted on NNs, our approach could show resiliency against fault injection and cache/memory attacks (see Section~\ref{sec:discussion} and Appendix~A for more information).

\section{Related Work}\label{Related Works}
\begin{table*}[t!]
\scriptsize
    \renewcommand{\arraystretch}{0.1}
        \begin{center}
        \caption{Summary of most recent side-channel attacks against DL accelerators.}\label{Tab_2}
            \begin{tabular}{|c|c|c|l|c|}
                \hline
                Paper&Targets&Side-channel Modality&\multicolumn{1}{c|}{Attack Scenario}&Implementation Platform\\
                \hline
                \makecell{Xiang et al.~\cite{xiang2020open}}&\makecell{DL Model Architecture}&Power&\makecell[l]{$\bullet$ Modeling the power consumption of different DL hardware accelerator components\\based on the number of additions and multiplications\\$\bullet$ Trained a classifier to reveal the DL Model architecture based on the captured power\\consumption traces}&\makecell{Raspberry Pi}\\
                \hline
                \makecell{DeepEM~\cite{yu2020deepem}}&\makecell{DL Model Architecture}&\makecell{EM}&\makecell[l]{$\bullet$ Presumption of a layer computations\\$\bullet$ Finding the number of parameters through each layer based on EM traces}&Pynq-Z1\\ 
                \hline
                \makecell{CSI NN~\cite{batina2019csi}}&\makecell{DL Model Architecture\\+Weights +AF}&Timing + EM&\makecell[l]{$\bullet$ Modeling all possible AF timing side-channel\\$\bullet$ Extracting the AF used in the DL Model Architecture\\$\bullet$ Distinguishing the EM patterns to find the number of layers and neurons\\$\bullet$ Launching CPA to reveal the weights}&\makecell{ARM Cortex-M3\\+ Atmel ATmega328P}\\
                \hline
                \makecell{Dubey et al.~\cite{dubey2020maskednet}}&\makecell{DL Model Weights}&Power&\makecell[l]{$\bullet$ Capturing the power consumption traces from changing status of pipeline registers\\$\bullet$ Launching a CPA based attack to reveal weights}&SAKURA-X FPGA board\\
                \hline
                \makecell{Yoshida et al.~\cite{yoshida2020model}}&\makecell{DL Model Weights}&Power&\makecell[l]{$\bullet$ Launching a CPA based attack to reveal weights}&Xilinx Spartan3-A\\
                \hline
            \end{tabular}
    \end{center}
    \vspace{-12pt}
\end{table*}
\begin{table*}[t!]
\scriptsize
    \setlength{\tabcolsep}{0.2em}
    \renewcommand{\arraystretch}{0.1}
        \begin{center}
        \caption{Summary of garbled DL accelerators and their features.}\label{Tab_1}
            \begin{tabular}{|c|c|c|l|c|}
                \hline
                Paper&Adversary Model&Approach&\multicolumn{1}{c|}{Contribution}&Implementatio Platform\\
                \hline
                \makecell{DeepSecure~\cite{rouhani2018deepsecure}}&HbC&Garbling&\makecell[l]{$\bullet$ Presentation of pre-processing approach\\
                $\bullet$ pre-processing step would reveal some information about the network parameters and structure of data cf.~\cite{riazi2019xonn}}&Intel Core i7 CPUs\\
                \hline
                \makecell{Chameleon~\cite{riazi2018chameleon}}&HbC&Hybrid&\makecell[l]{$\bullet$ Performs linear operations using additive secret sharing and nonlinear operations using Yao’s Garbled Circuits}&8-Core AMD CPU 3.7GHz\\
                \hline
                \makecell{Ball et al.~\cite{ball2019garbled}}&HbC&Hybrid&\makecell[l]{$\bullet$ Improvement of the BMR scheme~\cite{ball2016garbling} to support  Non-linear operations}&Intel Core i7-4790 CPUs\\
                \hline
                \makecell{XONN~\cite{riazi2019xonn}}&HbC&Garbling&\makecell[l]{$\bullet$ Support Binary NNs\\$\bullet$ Conversion of Matrix Multiplication to XNOR PopCount}&Intel Xeon CPU E5-2650\\
                \hline
                \makecell{TinyGarble2~\cite{hussain2020tinygarble2}}&HbC + Malicious &Garbling&\makecell[l]{$\bullet$ Provision of protection against malicious adversary\\$\bullet$ Alleviation garbling memory cost}&Intel Xeon CPU E5-2650\\
                \hline\hline
                \makecell{GarbledCPU~\cite{songhori2016garbledcpu}}&HbC&Garbling&\makecell[l]{$\bullet$ Presentation of FPGA accelerator for GC evaluation}&Virtex-7 FPGA\\
                \hline
                \makecell{RedCrypt~\cite{rouhani2018redcrypt}}&HbC&Garbling&\makecell[l]{$\bullet$ Minimizing the hardware architecture idle cycles to achieve scalable garbling}&Virtex UltraSCALE VCU108\\ 
                \hline
            \end{tabular}
    \end{center}
    \vspace{-12pt}
\end{table*}
\subsection{SCA against NNs}\label{Side-channel Based Attacks}
The main goals of side-channel attacks targeting DL hardware accelerators can be: (i) extraction of deep learning model (DL Model) architectures, and (ii) revealing NN parameters (i.e., weights and biases).
Table~\ref{Tab_2} summarizes some of these attacks.
In more details, Xiang et al.~\cite{xiang2020open} presented a power side-channel attack to extract the DL Model architecture.
For this purpose, they modeled the power consumption of different DL Model components such as pooling, activation functions, and convolutional layers based on how many addition and multiplication those components have.
Using these power consumption models, an SVM-based classifier was trained to reveal the DL Model architecture based on the captured power consumption traces when the DL Model runs on the hardware accelerator.
This line of research has also been pursued by Batina et al.~\cite{batina2019csi} who introduced an attack scenario based on the EM and timing side-channel to extract the number of layers, the number of neurons in each layer, weights, and activation functions (AF).
First, they modeled the timing side-channel of all possible AF (e.g., Relu or Tanh) and extracted the AF used in the NN by comparing the response time of the DL hardware accelerator when it executed the AF and the timing model of each possible AF.
This is followed by analyzing EM traces captured when the DL hardware accelerator runs, where the EM patterns determine the number of layers and number of neurons in each layer.
By feeding different random inputs to the accelerator and  capturing the EM traces, it was made possible to launch a Correlation
Power Analysis (CPA) to reveal the weights.
In another approach, Breier et al.~\cite{breier2021sniff} present a reverse engineering attack to extract the DL model weights and biases (parameters) with the help of fault injection on the last hidden layer of the network.

\vspace{-10pt}
\subsection{Security-preserving DL Accelerators}\label{Side-channel Protection Countermeasures}
To protect NNs against SCA, Liu et al.~\cite{liu2019mitigating} introduced a shuffling and fake memory-based approach to mitigate reverse engineering attacks that increase the run time of a DL hardware accelerator when the depth of the NN increases. 
Regarding the similarity between SCA launched against cryptographic implementations and DL accelerators, in a series of work, Dubey et al. have proposed hiding and masking techniques to protect NNs~\cite{dubey2020bomanet,dubey2020maskednet,dubey2022modulonet}. 
Yet, the differences between these implementations make the adaptation of known side-channel defenses challenging; for instance, integer arithmetic used in neural network computations that is different from modular arithmetic in cryptography, which has been addressed in~\cite{dubey2020maskednet,dubey2022modulonet}. 
Despite the impressive achievements presented in these studies, the approaches suffer from the known limitations of masking, i.e., their restriction to a specific side-channel security order. 
Furthermore, the implementation of masked DL models (i.e., a new circuit should be designed/implemented for different NNs) would be a challenging task. 
Moreover, masking cannot protect the architecture of DL accelerators. 

\begin{table}[t!]
\scriptsize
    \setlength{\tabcolsep}{0.5em}
    \renewcommand{\arraystretch}{0.5}
        \begin{center}
        \caption{State-of-the-art approaches vs. HWGN\textsuperscript{2} \small{(``P''arameters Secrecy of DL Model. ``U''pgradeable to/supporting malicious security model.  ``A''rchitecture protection of DL model. ``C''onstant round complexity. ``I''ndependence of a secondary server). Inspired by~\cite{riazi2019xonn}.} \label{Tab_10}}
            \begin{tabular}{|c|c|c|c|c|c|}
                \hline
                \textbf{Approach}&\textbf{P}&\textbf{U}&\textbf{A}&\textbf{C}&\textbf{I}\\ 
                
                \hline
                DeepSecure~\cite{rouhani2018deepsecure} &\textcolor{mygreen}{\checkmark}&\textcolor{mygreen}{\checkmark}&\textcolor{red}{\ding{55}}&\color{mygreen}\checkmark&\color{mygreen}\checkmark\\
                \hline 
                Chameleon~\cite{riazi2018chameleon} &\textcolor{mygreen}{\checkmark}&\textcolor{red}{\ding{55}}&\textcolor{red}{\ding{55}}&\textcolor{red}{\ding{55}}&\textcolor{red}{\ding{55}}\\
                \hline 
                XONN~\cite{riazi2019xonn} &\textcolor{mygreen}{\checkmark}&\textcolor{mygreen}{\checkmark}&\textcolor{red}{\ding{55}}&\color{mygreen}\checkmark&\color{mygreen}\checkmark\\
                \hline 
                BoMaNET~\cite{dubey2020bomanet}&\textcolor{mygreen}{\checkmark}&\textcolor{red}{\ding{55}}&\textcolor{red}{\ding{55}}&\color{mygreen}\checkmark&\color{mygreen}\checkmark\\ 
                \hline
                ModuloNET~\cite{dubey2022modulonet}&\textcolor{mygreen}{\checkmark}&\textcolor{red}{\ding{55}}&\textcolor{red}{\ding{55}}&\color{mygreen}\checkmark&\color{mygreen}\checkmark\\
                \hline
             TinyGarble2~\cite{hussain2020tinygarble2} &\textcolor{mygreen}{\checkmark}&\textcolor{mygreen}{\checkmark}&\textcolor{red}{\ding{55}}&\color{mygreen}\checkmark&\color{mygreen}\checkmark\\
                \hline\hline
                \makecell{RedCrypt~\cite{rouhani2018redcrypt}}&\color{mygreen}\checkmark&\textcolor{mygreen}{\checkmark}&\textcolor{red}{\ding{55}}&\color{mygreen}\checkmark&\textcolor{red}{\ding{55}}\\ 
                \hline
                \textbf{HWGN\textsuperscript{2} [This paper]} &\textcolor{mygreen}{\checkmark}&\textcolor{mygreen}{\checkmark}&\textcolor{mygreen}{\checkmark}&\color{mygreen}\checkmark&\color{mygreen}\checkmark\\
                \hline
            \end{tabular}
    \end{center}
    \vspace{-5pt}
\end{table}

\subsubsection{Garbled Accelerators}
Table~\ref{Tab_1} summarizes the most recent frameworks developed with regard to principles of SFE, more specifically, garbling. 
Among those proposals, GarbledCPU~\cite{songhori2016garbledcpu} and  RedCrypt~\cite{rouhani2018redcrypt} are of great importance to our work since they consider a hardware implementation of garbled circuits, whereas other relevant studies such as~\cite{songhori2015tinygarble,ball2016garbling,rouhani2018deepsecure,riazi2018chameleon,riazi2019xonn,hussain2020tinygarble2} devoted to software-based garbling engine/evaluator. 

\noindent\textbf{General-purpose hardware accelerators: }
In contrast to that,~\cite{songhori2016garbledcpu} has demonstrated a hardware garbling evaluator implemented on general-purpose sequential processors, where the privacy of NN architectures is also ensured. 
While benefiting from the simplicity of programming a processor, their design is specific to Microprocessor without Interlocked Pipelined Stages (MIPS) architecture. 
This has been addressed by introducing ARM2GC framework, where the circuit to be garbled/evaluated is the synthesized ARM processor circuit that can support pervasiveness and conditional execution~\cite{songhori2019arm2gc}. 
The efficiency in terms of hardware resources and communication cost has been reported as well. 
This is followed by introducing an FPGA-based garbling engine (FASE) that can be further extended to evaluator~\cite{hussain2019fase}. 
FASE is optimized to enable cloud servers to provide secure services to a large number of clients simultaneously without violating the privacy of their data; however, the privacy of circuits (e.g., NN architecture) is not considered in this study. 

\noindent\textbf{Hardware DL accelerators:}
RedCrypt attempts to enable cloud servers to provide high-throughput and power-efficient services to their clients in a real-time manner~\cite{rouhani2018redcrypt}. 
For this, FPGA platforms (Virtex UltraSCALE VCU108) have been used as a garbling core to present an efficient GC architecture with precise gate-level control per clock cycle, which ensures minimal idle cycles. 
This results in a multiple-fold improvement in the throughput of garbling operation compared to the previous hardware garbled circuit accelerator~\cite{songhori2016garbledcpu,songhori2015tinygarble}. 
In their scenario, a host CPU is involved in an OT to communicate the evaluator labels/input with the client, which may need high bandwidth. 
Although RedCrypt~\cite{rouhani2018redcrypt} has achieved significant improvement in computational efficiency, the DL model implemented on the FPGA cannot be easily diversified. 
Their proposed hardware DL accelerator suits a specific type of DL model and is built on the assumption that the network architecture is publicly available, which allows an adversary to launch an SCA attack easier~\cite{batina2019csi}. 
These shortcomings are tackled by HWGN\textsuperscript{2} that is NN-agnostic and guarantees the privacy of the DL model, i.e., the secrecy of its architecture. 
A qualitative comparison between the most state-of-the-art approaches and HWGN\textsuperscript{2} is provided in Table~\ref{Tab_10}. 
HWGN\textsuperscript{2} shares similarities with TinyGarble2~\cite{hussain2020tinygarble2}, although it is a software accelerator, whereas HWGN\textsuperscript{2} is a hardware one.

\section{Background}\label{sec:background}
\subsection{SFE/PFE Protocols}\label{SFE/PFE Protocol}
SFE protocols enable a group of participants to compute the correct output of some agreed-upon function $f$ applied to their secure inputs without revealing anything else. 
One of the commonly-applied SFE protocols is Yao's garbled circuit~\cite{yao1986generate}, a two-party computation protocol. To formalize this protocol, we employ the notions and definitions provided in~\cite{bellare2012foundations} to support modular and simple but effective analyses. 
In this regard, a garbling algorithm $Gb$ is a randomized algorithm, i.e., involves a degree of randomness. 
$Gb(f)$ is a triple of functions $(F, e, d) \leftarrow Gb(f)$ that accepts the function $f: \{0, 1\}^n \rightarrow \{0, 1\}^m$ and the security parameter $k$. 
$Gb(f)$ exhibits the following properties. 
The encoding function $e$ converts an initial input $x \in \{0, 1\}^n$ into a garbled input $X = e(x)$, which is given to the function $F$ to generate the garbled output $Y = F(X)$. 
In this regard, $e$ encodes a list of tokens (so-called labels), i.e., one pair for each bit in $x \in \{0, 1\}^n$:  $En(e, \cdot)$ uses the bits of $x = x_1 \cdots x_n$ to select from $e = (X^1_0 ,X^1_1, \cdots ,X^0_n,X^1_n)$ and obtain the sub-vector $X = X^{x_1}_1,\cdots,X^{x_n}_n$. 
By reversing this process, the decoding function $d$ generates the final output $y = d(Y)$, which must be equal to $f(x)$. 
In other words, $f$ is a combination of probabilistic functions $d\circ F\circ e$. 
More precisely, the garbling scheme $G=(Gb, En, De, Ev, ev)$ is composed of five algorithms as shown in Figure~\ref{fig:garbling_scheme}, where the strings $d$, $e$, $f$, and $F$ are used by the functions $De$, $En$, $ev$, and $Ev$ \textcolor{black}{(see Section~\ref{sec:foundation} for a concrete protocol flow in the case of NNs). }

\vspace{0.5ex}\noindent\textbf{Security of garbling schemes: } 
For a given scheme, the security can be roughly defined as the impossibility of acquiring any information beyond the final output $y$ if the party has access to $(F,X, d)$. 
Formally, this notion is explained by defining the side-information function $\varPhi(\cdot)$. 
Based on the definition of this function, an adversary cannot extract any information besides $y$ and $\varPhi(f)$ when the tuple $(F,X, d)$ is accessible. 
As an example of how the function $\varPhi(\cdot)$ is determined, note that for an SFE protocol, where the privacy of the function $f$ is not ensured, $\varPhi(f)=f$. 
Thus, the only thing that leaks is the function itself.
On the other hand, when a PFE protocol is run, $\varPhi(f)$ is the circuit/function's size, e.g., number of gates.

\vspace{0.5ex}\noindent\textbf{Oblivious transfer (OT):} 
This is a two party protocol where party~2 transfers some information to party~1 (so-called evaluator); however, party~2 remains oblivious to what information party~1 actually obtains. 
A form of OT widely used in various applications is known as ``chosen one-out-of-two'', denoted by  1-out-of-2 OT. 
In this case, party~2 has bits $X^0$ and $X^1$, and party~1 uses one private input bit $s$. 
After running the protocol, party~1 only gets the bit $X^s$, whereas party~2 does not obtain any information on the value of $s$, i.e., party~2 does not know which bit has been selected by party~1. 
This protocol can be extended to support the $n$-bit case, where party~1 bits $x_1, \cdots, x_n$ are applied to the input of party~2 $X^0_1,X^1_1, \cdots, X^0_n,X^1_n$ to obtain $X^{x_1}_1,\cdots,X^{x_n}_n$. 
This is possible by sequential repetition of the basic protocol~\cite{bellare2012foundations}. 
It has been proven that 1-out-of-2 OT is universal for 2-party SFE, i.e., OT schemes can be the main building block of SFE protocols~\cite{kilian1988founding}. 
\vspace{-10pt}
\subsection{Neural Network}\label{Neural Networks}
Neural Network is one of the main categories of machine learning, which refers to learning a non-linear function through multiple layers of neurons with the goal of predicting the output corresponding to a given input fed to a network trained on a dataset.
To perform such prediction, the input is fed to the first layer of the network (so-called \textit{input layer}), whereas in the next layers (so-called \textit{hidden layers}) the abstraction of the data takes place.
For a \emph{multi-layer perceptron} (MLP) that is a fully connected NN, each layer's input (including the input layer) is multiplied by neuron weights, added to the bias, and finally given to a commonly-applied \textit{activation functions} at the output of each layer (excluding the input layer).
The most common activation functions are Sigmoid, Tanh, and Rectified Linear Unit (ReLu).
The activation functions that might be used in DL models include linear, Sigmoid, and softmax. 

%% file: 3.Theory.tex
\section{Foundations of HWGN\textsuperscript{2} }\label{sec:foundation}

\noindent\textbf{Protocol flow:}
Here we provide insight into how SFE/PFE schemes can be tailored to the needs of a secure and private DL accelerator. 
According to the general flow illustrated in Figure~\ref{fig:garbling_scheme}, the goal of a garbling protocol $G$ is to evaluate a function $f$ against some inputs $x$ to obtain the output $y$.
The evaluator (i.e., the attacker) is never in possession of the raw NN binaries. 
Let $f=f_{NN}$ denote the function corresponding to the NN. 
The attacker aims to obtain the information on $f_{NN}$ by collecting the side-channel traces. 
To achieve this, here we give an example of SFE protocol $G$ that has OT at its core and follows Yao's garbling principle, i.e., the garbling protocol $G=(Gb, En, De, Ev, ev)$ as shown in Figure~\ref{fig:garbling_scheme}. 
To execute the protocol, the designer of the NN accelerator (garbler) conducts $(F, e, d) \leftarrow Gb(1^k,f)$ on inputs $1^k$ and $f$ and parses $(X^1_0 ,X^1_1, \cdots ,X^0_n,X^1_n) \leftarrow e$. 
Afterward, the garbler sends $F$ to the evaluator, i.e., the attacker. 
In order to perform the function $Ev$, the attacker and the garbler run the OT, where the former has the selection string $x$ and the latter party has already parsed $(X^1_0 ,X^1_1, \cdots ,X^0_n,X^1_n)$. 
Hence, the evaluator can obtain $X=X^{x_1}_1,\cdots,X^{x_n}_n$ and consequently, $y \leftarrow De(d, Ev(F,X))$. 
Note that even with the tuple $(F,X, d)$ in hand, the attacker cannot extract any information besides $y$ and $\varPhi(f)$.  
Moreover, although the NN provider has access to $(F,e,d)$, no information on $x$ leaks. 
In an inference scenario, $x$ represents the evaluator's input data. 
Nevertheless, if $G$ is an SFE scheme, $\varPhi(f)=f$. 

To construct a PFE scheme protecting the architecture, parameters, and hyperparameters of the NN that relies on the scheme $G$, we first define a  polynomial algorithm $\Pi$ that accepts the security parameter $k$ and the (private) input of the party~\cite{bellare2012foundations}. 
The PFE scheme is a pair $\mathcal{F} = (\Pi, ev)$, where $ev$ is as defined for the garbling scheme (see Section~\ref{IAS} for more information about $\Pi$). 
The scheme $\mathcal{F}$ enable us to securely compute the \emph{class} of functions $\{ev(f, \cdot) : f \in \{0, 1\}^*\}$, i.e., any function that $G$ can garble. 
The security of the PFE scheme $\mathcal{F}$ relies on the security of the SFE protocol underlying $\mathcal{F}$ (see Section~\ref{SFE/PFE Protocol}); however, $\varPhi(f)$ is the circuit size, i.e., the function $f$ remains private when executing the SFE protocol. 
In other words, the NN, its architecture, parameters and hyperparameters are now kept private from the attacker.

\noindent\textbf{Oblivious Inference: }
Oblivious inference tackles the problem of running the DL model on the user's input without revealing the input or the result to the other party (i.e., garbler in our case). 
For the latter, another interesting characteristic of SFE/PFE schemes is their ability to adapt to specific scenarios, where the output $y$ should also be protected. 
This would not be interesting in our case, where the security of the NN against SCA mounted by the evaluator is the objective. 
Nonetheless, for the sake of completeness, if the decryption of $Y$ should be performed securely, the privacy of inference results can easily be preserved by applying a one-time message authentication code (MAC) to the output and XORing the result with a random input to hide the outcome. 
These operations can be included in the design of the NN and naturally increase its size and the input fed by the garbler; however, the increase is linear in the number of output bits and considered inexpensive~\cite{lindell2007efficient}.

%% file: 3_1.Implementation.tex
\begin{figure}[t]
\centering \noindent
\includegraphics[width=0.85\columnwidth]{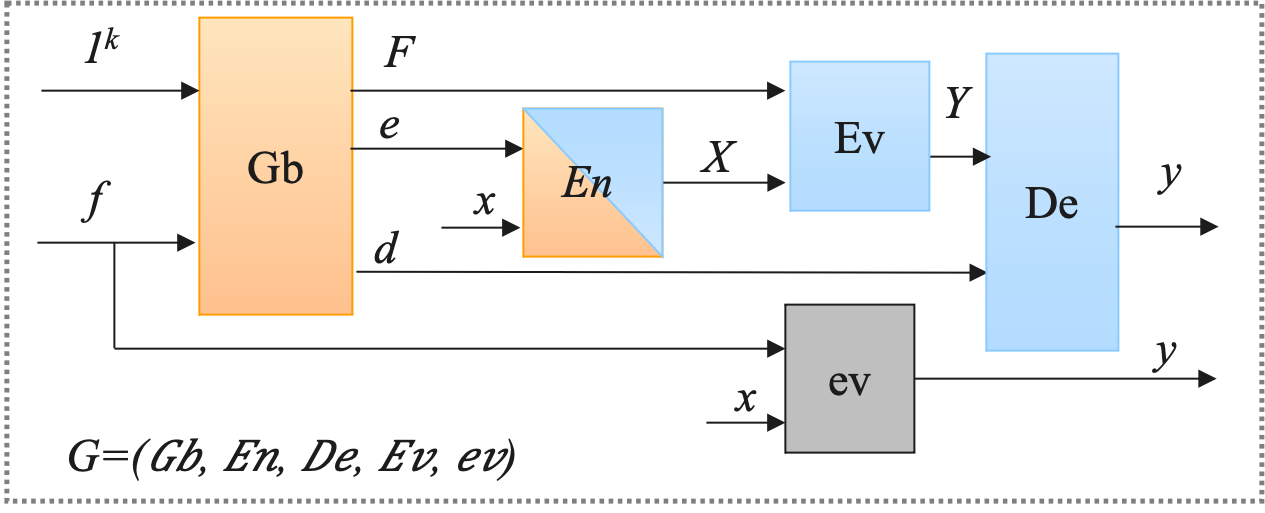}
\caption {A generic garbling scheme $G = (Gb, En, De, Ev, ev)$ cf.~\cite{bellare2012foundations}. Our proposed secure and private DL accelerator is built upon $G$. 
For HWGN\textsuperscript{2}, the blocks in orange show the operations performed by the NN vendor, whereas the blacks ones indicate the evaluator operations. 
$ev$ denotes the typical, unprotected evaluation of the function $f$ against the input $x$. 
}
\label{fig:garbling_scheme}
\end{figure}

\begin{figure}[t!]
\centering \noindent
\includegraphics[width=0.8\columnwidth]{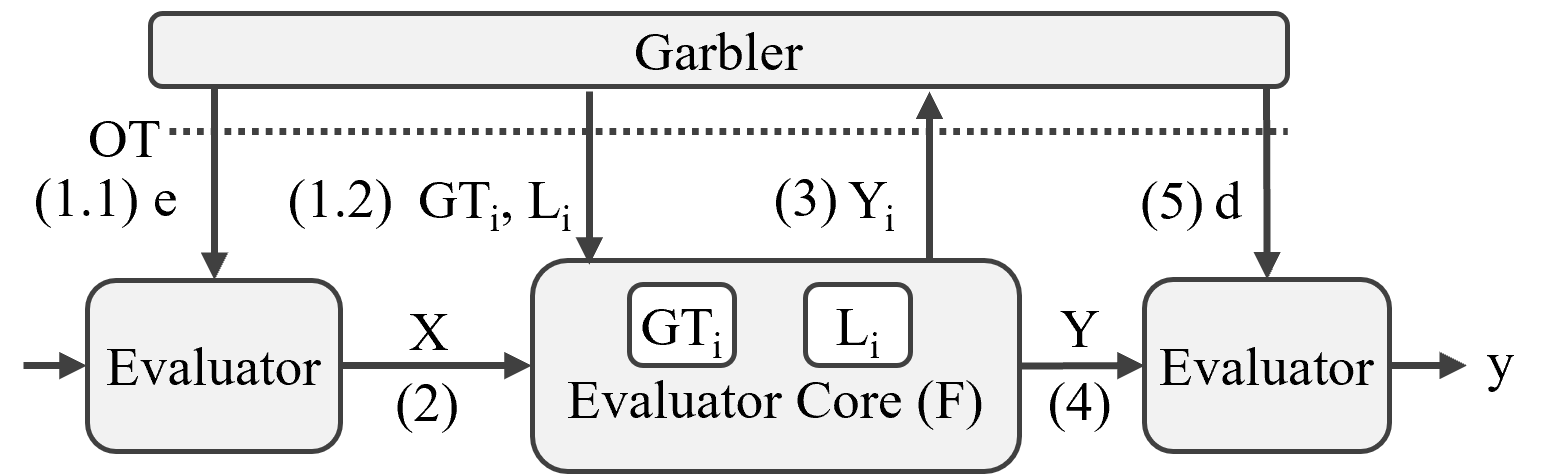}
\caption{Flow of HWGN\textsuperscript{2} ($L$: garbled wire labels, $GT$: garbled tables, $e$ and $d$: encryption and decryption labels, $X$: evaluator's garbled input, $Y$: garbled output, $Y_i$, $X_i$, $GT_i$, $L_i$: garbled input, output, tables, wire labels corresponding to $i^{th}$ sub-netlist, respectively, $y$: evaluator's raw output).}
\label{fig:HBC_flow2}
\end{figure}

\vspace{-10pt}
\subsection{Implementation of HWGN\textsuperscript{2}}\label{sec:improved_tinygarble2}
When defining the PFE scheme $\mathcal{F}$, it is mentioned that $\mathcal{F}$ can securely and privately compute \emph{any} function, which can be garbled by running the garbling scheme $G$. 
\acsac{Our garbled universal circuit $\mathcal{F}$ depends on the fact that a universal circuit is similar to a universal Turing machine~\cite{herken1988universal}, which can be realized by a general purpose processor cf.~\cite{songhori2015tinygarble,songhori2016garbledcpu}. } 
Note the difference between our goal, i.e., realizing $\mathcal{F}$, and one achieved in~\cite{wang2016secure}: optimizing the emulation of an entire \emph{public} MIPS program. 
Although we implemented a MIPS-based scheme, the prototypes can be extended to ARM processors. 
HWGN\textsuperscript{2} garbles the MIPS instruction set with a minimized memory and logical hardware resource utilization (see Section~\ref{Garbled_MIPS}). 

\noindent\textbf{Similarities between HWGN\textsuperscript{2} and TinyGarble2: }
One of the state-of-the-art GC frameworks is TinyGarble2~\cite{hussain2020tinygarble2} offering solely \emph{software} DL inference, without ensuring the privacy of the NN.  
HWGN\textsuperscript{2} remedies these shortcomings; however, it shares  similarities with TinyGarble2, namely regarding the flow of the protocol. 
The technique presented in TinyGarble2 is based on the division of a large netlist, such as DL models, into $i$ smaller sub-netlists and evaluating them one after another.
The size of the sub-netlists could be either one gate or equal to the total number of gates in the $f$ netlist.
The fewer gates included in each sub-netlist, the less memory utilization the gates require to be evaluated. 

Figure~\ref{fig:HBC_flow2} illustrates the flow of HWGN\textsuperscript{2} in the presence of an HbC adversary (for the extension to the malicious security model, see Appendix~A). 
First, the garbler chooses input encryption labels ($e$)  (Step~1.1). 
Afterward, instead of sending the complete set of GTs and L to the evaluator, in each cycle the garbler sends the evaluator a subset $GT_i$, $L_i$ (Step~1.2), and either $e$ (if the sub-netlist includes the gate with the inputs connected to the $f$ netlist) or $X_i$ (the garbled input corresponding to the sub-netlist). 
These subsets can be prepared offline and independent from the input of the evaluator. 
The evaluator also garbles her inputs as shown in Step~2, which is done offline as well.
In the next step, the evaluator evaluates the gate and sends the garbler the garbled output $Y_i$, i.e., garbled output of the $i^{th}$ sub-netlist (Step~3). 
This process repeats until all $i$ sub-netlists, excluding the gates whose output is connected to the NN outputs, are evaluated. 
Then in Step~4, the garbler sends the garbled tables and labels related to the gates that are connected to the NN output (so-called NN output layer).
After the evaluator evaluates all output layer-related gates, the garbler sends the decryption label (Step~5) along with the concatenated garbled outputs to the evaluator.
Finally, the evaluator decrypts the concatenated garbled output $Y$ and achieves his raw output $y$.
Also, instead of sending the complete set of GTs, L, and $e$ through one OT interaction, TinyGarble2 requires one OT interaction per sub-netlist.
The trade-off of minimizing the memory utilization using TinyGarble2 is the communication cost. 

\vspace{5pt}
\acsac{\noindent\textbf{What makes HWGN\textsuperscript{2} superior: }
Since HWGN\textsuperscript{2} is a hardware accelerator, it can benefit from parallelism offered by hardware platforms, e.g., FPGAs. 
HWGN\textsuperscript{2}, contrary to the previous software and hardware accelerators including TinyGarble family~\cite{songhori2015tinygarble,hussain2020tinygarble2}, also gives the flexibility of tuning the communication costs and hardware resource utilization to the garbler (e.g., NN provider).
In the applications where communication cost poses a limitation (such as real-time applications), one can implement DL hardware accelerators by sending the complete set of GTs, L, and $e$ through one OT interaction. 
This minimizes the communication cost while hardware resources are utilized at the maximum amount. 
In contrast, in the application with the limitation of hardware resources, one can use the HWGN\textsuperscript{2} that implements DL hardware accelerators with the sub-netlist size of one or a small number of gates. 
This leads to a noticeable strength of the HWGN\textsuperscript{2}: the evaluator has access to neither the netlist (GC mapping file, often comes in Simple Circuit Description, SCD, format~\cite{songhori2015tinygarble}) nor all GT and labels at once. 
Furthermore, scheduling is taken care of on the garbler-side; hence, the scheduler module is not accessible by the evaluator; hence, HWGN\textsuperscript{2} can be used to extend our framework to malicious security model and multiple-execution setting (for more details, see Section~\ref{sec:discussion} and Appendix~A). 

As opposed to TinyGarble2, HWGN\textsuperscript{2} implementation is based on the garbled MIPS architecture, making the circuit private (i.e., no information about the NN architecture leaks) as explained next. }

\begin{figure}[t]
\centering \noindent
\includegraphics[width=0.95\columnwidth]{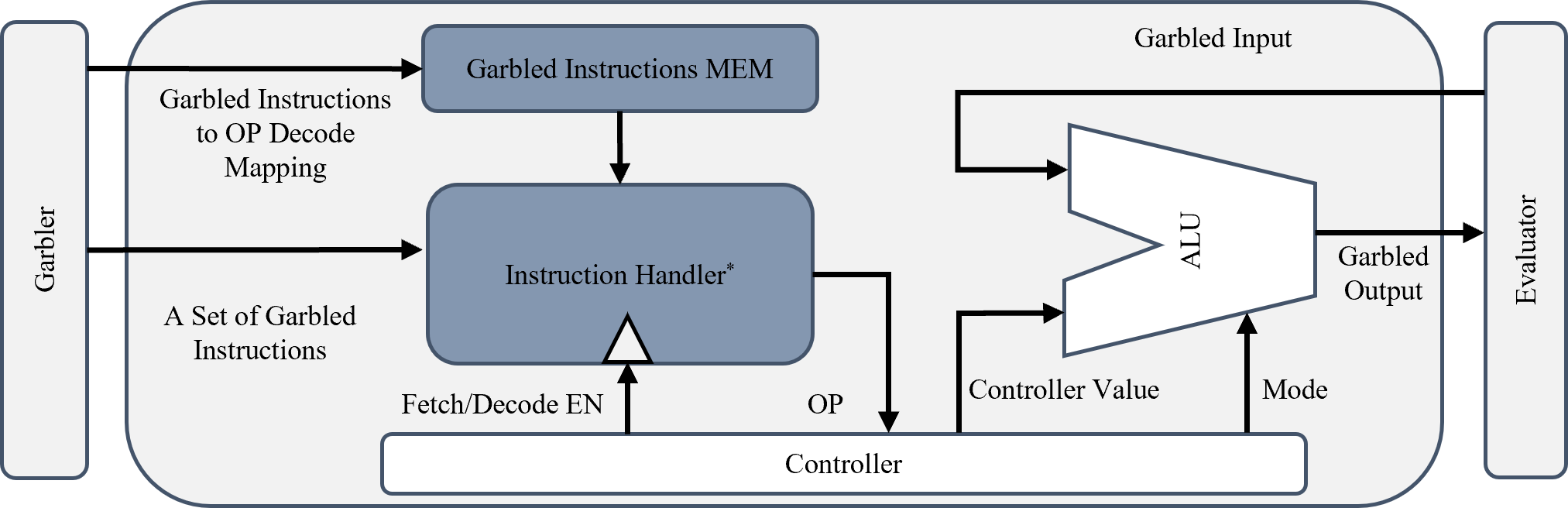}
\caption{Garbled MIPS evaluator, able to process any given number of instructions instead of a determined number. 
\acsac{The black modules are extended and improved versions of memory and instruction handler in Lite\_MIPS architecture~\cite{songhori2015tinygarble}: the instruction handler prepares the controller sequence by comparing the garbled MIPS instructions and the OP mapping.
The controller runs the process sequence by generating the ALU mode and executing the read and write operations. }}
\label{fig:Garbled_MIPS_architecture}
\end{figure}
\vspace{-10pt}
\subsubsection{MIPS Evaluator in HWGN\textsuperscript{2}} \label{Garbled_MIPS}
As explained before, in order to ensure the privacy of the NNs, the Boolean function representing the NN (so-called netlist) is converted to a set of reduced instruction set computing (RISC) instruction set architecture (ISA) and evaluated on a core that executes the MIPS instructions~\cite{kane1988mips}. 
\acsac{It might be thought that a subset of instructions required to execute the NN is sufficient to be garbled in order to reduce the overhead; however, this could increase the probability of guessing which instructions are used and, consequently, violates the privacy of the NN.  }

To implement HWGN\textsuperscript{2} on an FPGA, we modify Plasma~\cite{Plasma} MIPS execution core emulating a RISC instruction set on the FPGA, to act as the garbled MIPS evaluator.
Figure~\ref{fig:Garbled_MIPS_architecture} illustrates the architecture of our garbled MIPS evaluator.
The garbled evaluator receives three inputs: (i) a set of garbled instructions, (ii) the mapping for the instruction handler to fetch/decode the garbled instructions, and (iii) the evaluator's garbled input.
The combination of the first and second ones (i,ii) is the set of garbled tables and labels described before. 
Our garbled MIPS evaluator can evaluate the garbled MIPS instructions in two modes: (a) by receiving only one instruction and the operation code (OP) mapping and its corresponding instruction each cycle, i.e., the garbled evaluator with the capacity of one instruction per OT interaction, or (b) by receiving the complete set of instructions and their corresponding OP mapping at once.
\acsac{To achieve the resource-efficient implementation (mode i), we have modified the Lite\_MIPS instruction handler module in a way that the memory size related to the received garbled instructions (not the OP code mapping) decreases from 128 cells to only one cell.
The controller is further enhanced by discarding the unnecessary scheduler, SCD storage memory and its parsing modules and tailoring the core to need of only one instruction conversion per OT.
Moreover, we include the erase state in the instruction MEM controller, which sets all memory blocks to 0 after converting each garbled instruction to the OP code.
\textcolor{black}{To take advantage of the resource-efficient implementation, an extra step should be taken to divide the netlist into the sub-netlists with the number of gates selected by the user. 
The sub-netlists are fed to HWGN\textsuperscript{2} in the same order provided in the SCD file.
This allows the user to make a trade-off between the resource efficiency and performance of the HWGN\textsuperscript{2}.}

Specifically, in the first mode (a), in the first step, the instruction handler module receives one garbled instruction and OP mapping (all possible combinations of garbled MIPS instructions necessary to follow SFE protocol), which are stored in the instructions memory (MEM).
In the next step, the instruction handler compares the given garbled instructions with garbled instructions MEM information and converts each garbled instruction to a set of OPs.
Finally, the instruction handler sends the OP to the arithmetic-logic unit (ALU), erase instructions MEM, and repeats above-mentioned steps for the next garbled instructions.
In the second mode (b), however, the instruction handler module works similarly to the Lite\_MIPS architecture cf.~\cite{songhori2015tinygarble}. }
As both instruction sets and decode mapping are garbled on the garbler side, the evaluator cannot decrypt the garbled instructions due to the lack of the decryption key.
Therefore, the garbler's inputs and the DL model parameters are secure following the SFE and PFE protocols.


%% file: 4.Practical_Implementation.tex
\section{Evaluation of HWGN\textsuperscript{2}} \label{IAS}
\subsection{Resource Utilization} \label{Implementation_Evaluation}
To understand the interplay between communication cost, hardware resources utilization, and performance, we have synthesized the garbled evaluator with the capacity of $1$ and $2345$ (complete set of instructions) garbled MIPS instructions per one OT interaction. 
We have used Xilinx Vivado 2021 to synthesize our design and generate a bittsream.
To make sure that the bitstream is correctly associated with the design, we have disable Xilinx Vivado place-and-route optimization and also utilize the DONT-TOUCH attribute.
\acsac{The garbling framework considered in our implementations is JustGarble~\cite{bellare2013efficient}, also embedded in TinyGarble2 framework~\cite{hussain2020tinygarble2}, which enjoys garbling optimization techniques such as Free-XOR~\cite{kolesnikov2008improved}, Row Reduction~\cite{naor1999privacy}, and Garbling with a Fixed-key Block Cipher~\cite{bellare2013efficient}. }
Our implementation is applied against three typical MLPs: the first one, with 784 neurons in its input layer, three hidden layers each with 1024 neurons, and an output layer with 10 neurons that is trained on MNIST (hereafter called \textbf{BM1}). 
The results for applying state-of-the-art approaches against BM1 have been presented in~\cite{rouhani2018redcrypt,dubey2022modulonet,songhori2016garbledcpu}. 
The second MLP, \textbf{BM2}, has 784, 5, 5, and 10 neuron in its input, 2 hidden, and output layers, respectively. 
The third MLP, \textbf{BM3}, consists of 784, 6, 5, 5, and 10 neurons in its input, 3 hidden, and output layers, respectively.

\begin{table}[t!]
\scriptsize
    \setlength{\tabcolsep}{0.4em}
    \renewcommand{\arraystretch}{0.1}
        \begin{center}
        \caption{Hardware resource utilization and OT cost comparison between approaches applied against BM1.}\label{Tab_3}
            \begin{tabular}{|c|c|c|c|}
                \hline
                Approach&LUT&FF&OT Interaction\\
                \hline
                \makecell{Plasma ~\cite{Plasma}}&$1773$&$1255$&N/A\\
                \hline
                \makecell{GarbledCPU ~\cite{songhori2016garbledcpu}}&$21229$&$22035$&$2$\\
                \hline
                
                \makecell{RedCrypt ~\cite{rouhani2018redcrypt} (One MAC Unit)}&$111000$&$84000$&$2$\\
                \hline
                
                \makecell{BoMaNET ~\cite{dubey2020bomanet}}&$9833$&$7624$&N/A\\
                \hline
                \makecell{ModulaNET ~\cite{dubey2022modulonet}}&$5635$&$5009$&N/A\\
                \hline
                \makecell{\textbf{HWGN\textsuperscript{2}} \textbf{(1 instruction per} \textbf{OT interaction)}}&\textbf{$1775$}&\textbf{$1278$}&\textbf{$2346$}\\
                \hline
            \end{tabular}
    \end{center}
   \vspace{-5pt}
\end{table}

Table~\ref{Tab_3} shows a comparison between the hardware utilization and OT cost of \acsac{an unprotected MIPS evaluator core (Plasma\cite{Plasma}), HWGN\textsuperscript{2} and the state-of-the-art approaches applied to BM1.
To give an insight into how much overhead cost the protection approaches impose, we have implemented Plasma core, an unprotected MIPS evaluator core on an Artix-7 FPGA. }
Note that we choose this architecture for the sake of a better comparison with the state-of-the-art solutions, e.g.,~\cite{dubey2022modulonet}. 
It is also worth mentioning that since the ultimate goal of our paper is to demonstrate the applicability of garbling techniques for side-channel resiliency, the network mentioned above is chosen to serve as a proof of concept. 
As the HWGN\textsuperscript{2} processes the garbled instructions and inputs with the width of 32-bits, to have a fair comparison, we include the 32-bit MAC unit~\cite{rouhani2018redcrypt} in the resource utilization reported in Table~\ref{Tab_3}. 

In Table~\ref{Tab_3}, BoMaNET and ModulaNET do not use OT to exchange their inputs.
RedCrypt uses two OT interactions, one for the evaluator's input and another for the evaluator's output. 
However, in HWGN\textsuperscript{2}, in addition to the input and output labels exchange OT requirement, HWGN\textsuperscript{2} requires $M$ more OT interactions, where $M$ is the number of sub-netlists.
\acsac{There is an important observation made from Table~\ref{Tab_3}: HWGN\textsuperscript{2} with the capacity of one instruction per OT interaction utilizes $0.0011\times$ and $0.018\times$ more logical and memory hardware resources, respectively, compared to an unprotected MIPS evaluator.
The reason behind this efficiency is the size of instruction memory which stores only one instruction per OT interaction instead of the complete set of instructions. }
As mentioned in Section~\ref{sec:improved_tinygarble2}, to minimize resource utilization, one should sacrifice the communication cost, leading to an increased execution time. 
Hence, we set the size of the sub-netlist to just one gate, and every four gates are converted to a garbled instruction: $M =N_{gate}/4$, where $N_{gate}$ is the number of gates in the netlist.
In this setting, HWGN\textsuperscript{2} requires $2 +9380/4 = 2346$ OT interactions, where $9380$ is the number of gates included in the BM1 netlist.
Based on the results of Table~\ref{Tab_3}, HWGN\textsuperscript{2} utilizes up to roughly $11\times$, $4.5\times$, $3.2\times$, and $62.5\times$ fewer logical and $16\times$, $6\times$, $4\times$, and $66\times$ less memory hardware resources compared to GarbledCPU~\cite{songhori2016garbledcpu}, BoMaNET~\cite{dubey2020bomanet}, ModuloNET~\cite{dubey2022modulonet}, and RedCrypt~\cite{rouhani2018redcrypt}, respectively. \textcolor{black}{Based on results presented in Table~\ref{Tab_3}, HWGN\textsuperscript{2} is the most resource-efficient approach compared to the state-of-the-art approaches.}
\begin{table}[t!]
\scriptsize
    \setlength{\tabcolsep}{0.4em}
    \renewcommand{\arraystretch}{0.1}
        \begin{center}
        \caption{BM1 implemented by HWGN\textsuperscript{2} with different capacities of instruction per OT interaction.}\label{Tab_4}
            \begin{tabular}{|c|c|c|c|}
                \hline
                \makecell{Instruction capacity per OT interaction}&LUT&FF&OT Interaction\\
                \hline
                \makecell{$1$}&$1775$&$1278$&$2346$\\
                \hline
                \textbf{Complete set of instructions}&\textbf{$94701$}&\textbf{$52534$}&\textbf{$2$}\\
                \hline
            \end{tabular}
    \end{center}
    \vspace{-5pt}
\end{table}
In real-time applications where the execution time is the bottleneck, the OT interactions must be minimum~\cite{rouhani2018redcrypt}.
Therefore, in Table~\ref{Tab_4}, we have reported the hardware resource utilization in two cases: (i) when the number of OT interactions is maximum (first row) and (ii) when the number of OT interactions is minimum (second row).
The results in Table~\ref{Tab_4} are for the implementation of BM1. 
\textcolor{black}{As shown in Table~\ref{Tab_6}, HWGN\textsuperscript{2} with the maximum performance is $2.5\times$ faster than GarbledCPU~\cite{songhori2016garbledcpu}. Performance of HWGN\textsuperscript{2} is close to the performance of Redcrypt~\cite{rouhani2018redcrypt}, the fastest state-of-the-art approach, while utilizing $62.5\times$ fewer logical and $66\times$ less memory than Redcrypt~\cite{rouhani2018redcrypt}.}

\begin{table}[t!]
\scriptsize
    \setlength{\tabcolsep}{0.4em}
    \renewcommand{\arraystretch}{0.1}
        \begin{center}
        \caption{Execution time and communication cost comparison between HWGN\textsuperscript{2} and the state-of-the-art approaches (for BM1). Results for~\cite{songhori2016garbledcpu} and HWGN\textsuperscript{2} are reported based on FPGA with clock frequency equals to $20MHZ$. (N/R: not reported, inst.: instructions).}\label{Tab_6}
            \begin{tabular}{|c|c|c|}
                \hline
                Approach&Time (Sec)&Comminucation (MB)\\
                \hline
                \makecell{GarbledCPU ~\cite{songhori2016garbledcpu}}&$1.74$&N/R\\
                \hline
                \makecell{RedCrypt~\cite{rouhani2018redcrypt}}&$0.63$&$5520$\\
                \hline
                \makecell{\textbf{HWGN\textsuperscript{2} (Complete set of inst. per OT interaction)}}&\textbf{$0.68$}&\textbf{$619$}\\
                \hline\hline
                \makecell{TinyGarble2~\cite{hussain2020tinygarble2}}&$9.1$&$7.16$\\
                \hline
                \makecell{\textbf{HWGN\textsuperscript{2}} \textbf{(1 inst. per} \textbf{OT interaction)}}&$3.25$&$12.39$\\
                \hline
            \end{tabular}
    \end{center}
    \vspace{-7pt}
\end{table}

\begin{table}[t!]
\scriptsize
    \setlength{\tabcolsep}{0.4em}
        \begin{center}
        \caption{Execution time and communication cost comparison between BM1 accelerator and XNOR-based version of that.}\label{Tab_7}
            \begin{tabular}{|c|c|c|c|c|}
                \hline
                Architecture&\#Instructions&OT Interaction&Execution Time (Sec)&Communication (MB)\\
                \hline
                \makecell{BM1}&$2345$&$2346$&$3.25$&$12.39$\\
                \hline
                XNOR-based BM1&$1629$&$1631$&$2.31$&$9.71$\\
                \hline
            \end{tabular}
    \end{center}
    \vspace{-5pt}
\end{table}

\subsubsection{Execution Time and Communication Cost Evaluation} \label{Execution_Time}
To evaluate the cost of HWGN\textsuperscript{2} in terms of execution time, we have used a machine with Intel Core i7-7700 CPU @ 3.60GHz, 16 GBs RAM, and Linux Ubuntu 20 as the garbler and an ARTIX\-7 FPGA board as the evaluator, which has a clock frequency of 20 MHz.
All the garbled instructions, their MEM values, and labels are generated offline and not included in the execution time.
To communicate with the FPGA, for the sake of comparison, we have used HostCPU presented in~\cite{rouhani2018redcrypt}. 
\acsac{Note that in a real-world application, where the communication is performed over high latency links, the protocol execution remains fast due to the constant number of rounds in Yao's GC underlying our design cf.~\cite{juvekar2018gazelle,lindell2019efficient}. }
Moreover, we have used the EMP-toolkit~\cite{wang2017faster} to establish the OT interaction between the garbler and the HostCPU.
Table~\ref{Tab_6} shows the execution time and communication cost comparison between HWGN\textsuperscript{2} and the state-of-the-art approaches employed against BM1. 
The memory footprint of classical GC approaches is $O(I+N_{gate})$, where $I$ is the number of input wires and $N_{gate}$ is the number of gates in the netlist.
In contrast, the memory footprint of HWGN\textsuperscript{2} and TinyGarble2 is the same: $O(I+N_{gate,m}+i_m)$ where $N_{gate,m}$ is the number of gates in the largest among sub-netlists included in the design, and $i_m$ is the number of inputs of the sub-netlist, which equals 1 and 2, respectively, in the case of HWGN\textsuperscript{2} with the instruction capacity 1 per OT interaction. 

To compare the execution time and communication cost of TinyGarbled2 with our approach, we have chosen the semi-honest mode when using their framework.
HWGN\textsuperscript{2} outperforms the TinyGarble2 implemented on CPU thanks to the parallel implementation made possible by the FPGA. 
On the other hand, when minimizing the OT interactions by investing more hardware resource utilization, HWGN\textsuperscript{2} has a performance close to the RedCrypt with $62.5\times$ fewer logical and $66\times$ less memory utilization.

As an optimization technique, we have implemented the XNOR-based BM1. 
As the XOR operation is free in the garbling protocol~\cite{kolesnikov2008improved}, it is possible to decrease the size of the garbled netlists, which results in fewer instructions to be executed.
Table~\ref{Tab_7} shows a comparison between two architectures. 
Using an XNOR-based implementation of a DL hardware accelerator decreases the number of instructions, leading to a less OT cost and execution time.
The only limitation of this optimization is that the weights of the DL model must be binarized, and such binarization may slightly decrease the DL hardware accelerator output accuracy (see Section~\ref{sec:discussion} for more details).
Nevertheless, there are methods devised to deal with this, which can be adopted to bring significant benefits to garbled DL accelerators in terms of both OT cost and execution time.

\subsection{Side-channel Evaluation} \label{Side-channel_eval}
\subsubsection{Side-channel Measurement Setup}\label{SC_measurement_setup}
HWGN\textsuperscript{2} has been implemented on Artix-7 FPGA device XC7AT100T with package number FTG256. 
We have captured the power and EM traces (see Appendix~C) using Riscure setup, including LeCroy wavePro 725Zi as the setup oscilloscope.
To prevent information loss, we have set our design frequency to $1.5 MHz$ and the oscilloscope sampling frequency to $127.5 MHz$ to reduce the time- and memory-complexity of the test. 
For each clock cycle, we have acquired $85$ sample points.
Decreasing the design frequency with the purpose of acquiring high-resolution side-channel traces made our design execution time $3.25$ to $4.73$ seconds for each classification performed by BM1.
For this network, acquiring side-channel traces in the the order of millions has high time complexity.
Therefore, similar to~\cite{dubey2022modulonet}, another MLP architecture, namely, \textbf{BM2} is used for traces collection. 
The changes in MLP architecture hyperparameters allowed us to execute each classification in $312 ms$ using HWGN\textsuperscript{2} with $1.5 MHz$ design frequency.
As HWGN\textsuperscript{2} executes each instruction separately in a sequential manner and the nature of the NNs is repetitive, we argue that the smaller MLP architecture can represent a larger one in terms of leakage.

\subsubsection{Leakage Evaluation}
We have used a common methodology, namely Test Vector Leakage Assessment (TVLA) test, to evaluate HWGN\textsuperscript{2} leakage resiliency. 
Although the TVLA test is subject to two disadvantages -- false positive/negative results and limited ability to reveal all points of interests~\cite{durvaux2016improved, moradi2018leakage, standaert2018not} -- it is still the most common methodology used in recent papers to evaluate the resiliency of the approach against side-channel leakage.

In the TVLA test methodology, Welch’s t-test is used to check the similarity between two trace groups captured from two populations of inputs.
Welch’s t-test calculates the t-score based on the following equation: $t=(\mu_1-\mu_2)/\sqrt{(s_1^2/n_1^2)+(s_2^2/n_2^2)}$, where $\mu_1$ and $\mu_2$ are the means, $s_1$ and $s_2$ are the standard deviations, and $n_1$ and $n_2$ are the total number of the captured traces for first and second population, respectively.
Based on the null-hypothesis, if two populations are chosen from one distribution, their corresponding t-score must be less that $\pm4.5$.
Exceeding t-score magnitude of $4.5$ (so-called null-hypothesis) means the design is subject to side-channel leakage with probability greater than $99.99\%$.
In our setup, we choose the non-specific fixed vs. random t-test in a way that our setup, first, captures the power consumption/EM traces from a fixed input computation for all the traces; then, the experiment repeats for a set of randomly generated inputs.
Based on the two captured traces, for fixed and random inputs, our setup calculates the t-score based on the aforementioned equation.

\subsubsection{Power Side-channel Leakage Assessment}\label{TVLA_Power}
\begin{figure}[t!]
\centering \noindent
\includegraphics[width=\columnwidth]{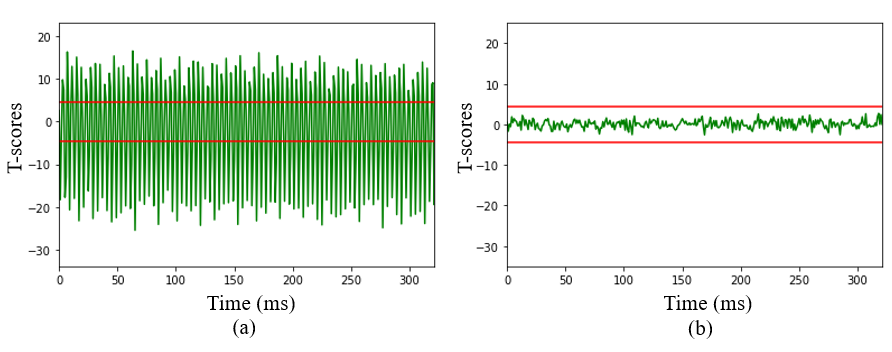}
\caption{TVLA test results for implementation of BM2 on (a) an unprotected MIPS core and (b) HWGN\textsuperscript{2} with the capacity of one instruction per OT (calculated for 10K traces).}
\label{fig:unprotected_garbled_comparison}
\end{figure}

To illustrate the side-channel protection offered by HWGN\textsuperscript{2}, we have mounted the TVLA test on power traces of an unprotected MIPS core, Plasma core presented by Opencores projects~\cite{Plasma}, and HWGN\textsuperscript{2}, with the capacity of one instruction per OT interaction.
Figure~\ref{fig:unprotected_garbled_comparison} shows the TVLA results of an unprotected MIPS core and HWGN\textsuperscript{2}, with the capacity of one instruction per OT interaction.
The t-scores are calculated based on $10000$ captured traces, $5000$ for each fixed and random input population.
As one can observe, an unprotected MIPS core t-score has exceeded the $\pm4.5$ threshold with only $10000$ traces, while the HWGN\textsuperscript{2}'s t-score remains below the threshold.

To have a design with leakage resiliency, the t-score results must remain below the threshold with the traces populations in the order of millions~\cite{de2018hardware,standaert2018not,moradi2018leakage,dubey2022modulonet}.
Hence, in the next experiment, we have captured a total of $2$ million ($2M$) traces, $1M$ traces for each fixed and random input populations.
A low t-score, less than $\pm4.5$, calculated from a trace population in the order of millions confirms the protection strength of HWGN\textsuperscript{2}.
It should be noted that these traces are captured in the low-noise setup (i.e., more optimistic for the attacker) while in the actual scenario, the number of traces to break the garbling scheme should be significantly higher due to more noisy environments. 

\begin{figure}[t!]
\centering \noindent
\includegraphics[width=\columnwidth]{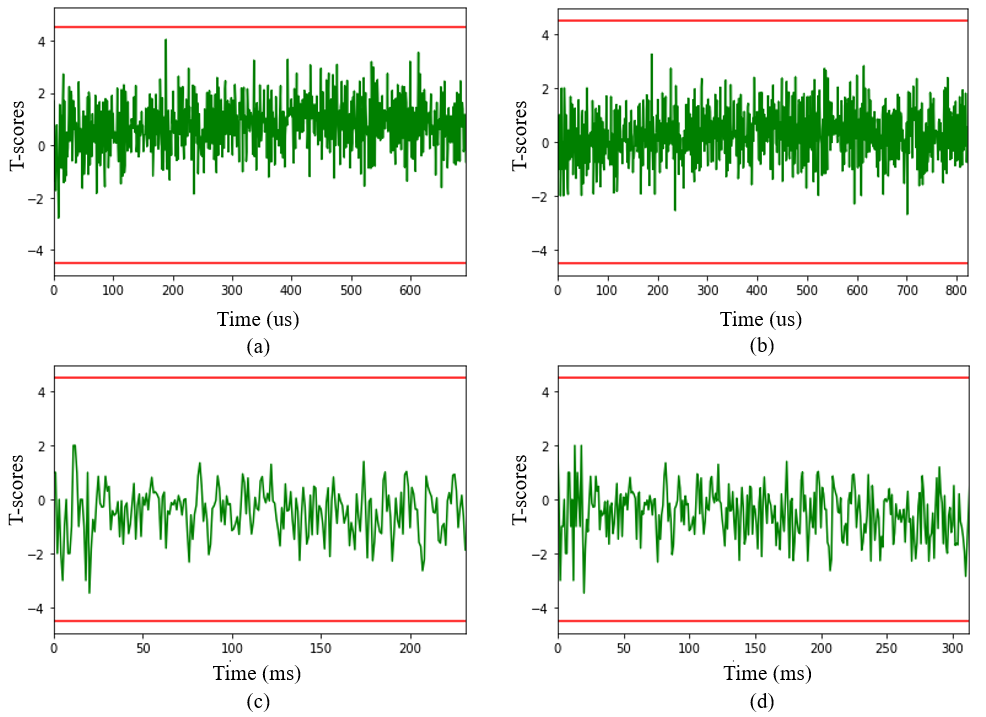}
\caption{TVLA test results (a) HWGN\textsuperscript{2} applied to XNOR-based BM2 (capacity whole set of instructions per OT) (b) BM2 with the capacity of complete set of instructions per OT interaction, (c) HWGN\textsuperscript{2} applied to XNOR-based BM2 (capacity 1 instruction per OT), and (d) BM2 with the capacity of 1 instruction per OT (calculated for $2M$ power traces). } 
\label{fig:garbled_capacity_comparison_2M}
\vspace{-3pt}
\end{figure}

As a proof of concept that HWGN\textsuperscript{2} side-channel resiliency is independent of the function or architecture we also mount the TVLA test on two more implementations: XNOR-based DL hardware accelerator and DL hardware accelerator.
Figure~\ref{fig:garbled_capacity_comparison_2M}, (a) and (c), illustrates the t-score of HWGN\textsuperscript{2} applied to XNOR-based BM2, with the capacity of complete set of instructions per OT interaction and one instruction per OT, respectively. 
As can be seen, the t-scores of HWGN\textsuperscript{2} stay below the threshold of $\pm4.5$ for different cases of instruction capacity per OT interaction and the function or architecture implemented on an FPGA using HWGN\textsuperscript{2}.
The t-scores in Figure~\ref{fig:garbled_capacity_comparison_2M}, (b) and (d), indicate that not only HWGN\textsuperscript{2} with the capacity of one instruction per OT interaction provides a strong protection against power side-channel attacks but also changes in the number of instruction capacity per OT interaction does not affect this protection (for results of EM SCA, see Appendix~C).


\begin{figure}[t!]
\centering \noindent
\includegraphics[width=\columnwidth]{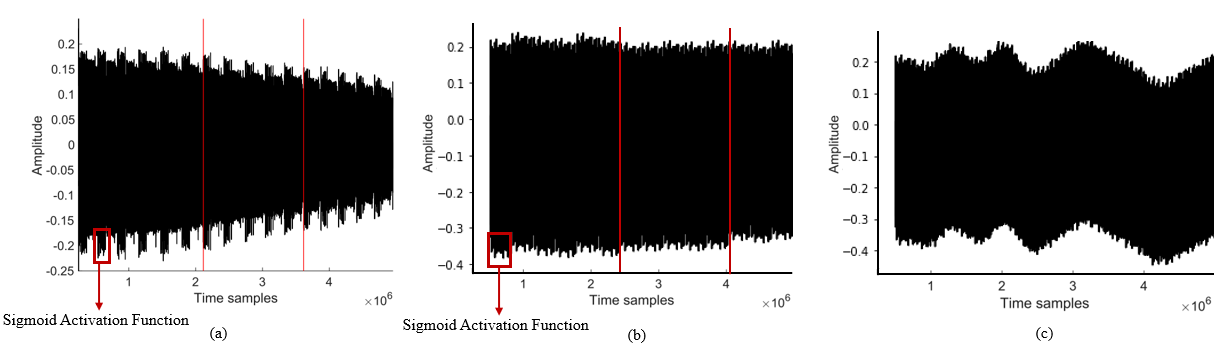}
\caption{A randomly chosen EM trace pattern captured from the implementation of BM3 on (a) Atmel ATmega328P microcontroller~\cite{batina2019csi} (b) FPGA with unprotected MIPS evaluator~\cite{Plasma} (c) with HWGN\textsuperscript{2}. Red lines correspond to time-samples, where the unprotected evaluators start the next layer evaluation. }
\label{fig:EM_comparison}
\vspace{-3pt}
\end{figure}

\noindent\textbf{Can we see the architecture-related patterns?} 
Based on the attack presented by Batina et al.~\cite{batina2019csi}, revealing the DL model architecture can enhance the attacker's ability to obtain DL model parameters. 
They showed that the EM trace captured from an unprotected DL model implementation on Atmel ATmega328P microcontroller, which follows the MIPS architecture same as HWGN\textsuperscript{2}, with three hidden layers containing 6, 5, and 5 neurons, respectively, has a pattern in which the number of layers and neurons can be revealed.
They have used LeCroy WaveRunner 610Zi oscilloscope and RF-U 5-2 near-field EM probe to capture EM traces.
To examine if we observe the same patterns as reported in~\cite{batina2019csi}, we have implemented the same DL model, \textbf{BM3}, and captured 100K EM traces. 
Figure~\ref{fig:EM_comparison}(a) illustrates the captured EM traces from an Atmel ATmega328P microcontroller taken from~\cite{batina2019csi}, whereas Figure~\ref{fig:EM_comparison}(b) show the traces collected from our unprotected MIPS evaluator core~\cite{Plasma}, and Figure~\ref{fig:EM_comparison}(c) presents the captured EM traces of HWGN\textsuperscript{2} for a randomly chosen EM trace.
From Figure~\ref{fig:EM_comparison}, it is observable that there exists a pattern, in which the number of the layers and neurons can be seen, similar to the observation made by Batina et al.~\cite{batina2019csi}: the red lines indicate the borders when MIPS evaluator starts the next hidden layer evaluation and the red squares correspond to the EM peak of Sigmoid AF evaluation.
In the case of the HWGN\textsuperscript{2}, EM traces do not follow a pattern which could result in revealing the DL model architecture.
The reason behind these irregular patterns is that each garbled instruction is encrypted; therefore, in the evaluation phase, the HWGN\textsuperscript{2} treats them as two nonidentical instructions, although the generated OP corresponding to them is the same.

%% file: 6.Discussion.tex
\section{Discussion}\label{sec:discussion}
\vspace{-2pt}
\noindent\textbf{Security in the presence of malicious adversary -- a byproduct.}
According to our adversary model defined in Section~\ref{Adversary Model}, the garbler may act as a malicious adversary to extract information related to the user's input by generating a corrupted GC~\cite{lindell2016fast}. 
Note that it does not match the scenario defined for SCA, although as a byproduct, one can consider the extension to the malicious security model. 
In doing so, to protect the evaluator from the malicious garbler, the concept of cut-and-choose, see, e.g.,~\cite{lindell2007efficient,lindell2016fast}, has been widely employed to ensure the security of the evaluator's input. 
The principal behind the cut-and-choose protocol is that the garbler constructs many GCs instead of one and sends them to the evaluator.
The evaluator then, chooses a random set of GCs, opens them, and checks whether the GCs are constructed correctly or not. 
To cope with the computation and communication overhead, Lindel has introduced a fast cut-and-choose protocol based on the majority outputs calculation with a cheating probability of at most $2^{-s}$ where $s$ is the number of constructed GCs by the garbler~\cite{lindell2016fast}, e.g., to achieve a cheating probability of at most $2^{-40}$, garbler must generate 40 GCs~\cite{pinkas2009secure,lindell2016fast}. 
It is remarkable that not every available garbling tool supports the malicious security model, although TinyGarbled2 can handle this, of course, with some modifications to satisfy the hardware implementation (e.g. pipelining). 
Therefore, if needed, HWGN\textsuperscript{2} can be easily upgraded to support the security under the malicious adversary scenario. 
To the best of our knowledge, HWGN\textsuperscript{2} is the first hardware accelerator to fulfill this need. 
The details of our implementation and results are given in Appendix~A. 

\vspace{4pt}\noindent\textbf{Multiple-execution setting.}
The multiple-execution setting has been understudied in the existing literature on the implementation of private DL accelerators. 
This is despite of the cryptographic protocols developed to guarantee the security of GCs under this circumstance. 
The multiple-execution setting is relevant if the DL accelerator (possibly with different inputs) is evaluated multiple times, either in parallel or sequentially.
When executing the protocol for the first time, it could be feasible to send the entire set of inputs at once (sometimes called parallel execution). 
Nonetheless, the sequential execution of the OT and the SFE protocol built upon leads to a more resource-efficient implementation since all the labels should not be stored together. 
Note that to guarantee security, a fresh garbled function $F$ for each user/a set of inputs should be generated. 
To extend the protocol to multiple-execution setting, a straightforward approach is to run a maliciously-secure protocol multiple times. 
In this regard, if the cheating probability is kept as before, i.e., at most $2^{-s}$, the number of constructed GCs should be increased to $s+\log t$ from $s$, where $t$ is the number of executions. 
This overhead can be indeed reduced if a more efficient protocol, e.g., one presented in~\cite{huang2014amortizing}, is considered. 
Noteworthy is the fact that, this protocol has been built upon the cut-and-choose protocol~\cite{lindell2016fast}, whose overhead for HW DL accelerators is discussed in Appendix~A.  
We argue that as the core computation for HWGN\textsuperscript{2} is the same in HbC, malicious, and multiple-execution settings, no side-channel leakage can be observed in the two latter cases either.

\vspace{4pt}\noindent\textbf{Trade-off between accuracy and efficiency.}
Thanks to the binary values and bitwise operations, Binarized NNs (BNNs) can perform the computation efficiently, although at the price of exhibiting less precision compared to their full precision counterparts. 
For instance, XNOR can perform the multiplication in a bitwise manner. 
Afterward, the accumulation needed by the dot product can be done by counting the number of bits set to ``1'' in a group of XNOR products, multiplying this value by 2 and subtracting the total number of bits producing an integer value. 
Alternatively, a Popcount instruction can be used from the processor instruction sets. 
Compared to multi-bit floating-point or fixed-point multiplication and accumulation, these bitwise operations are much faster and hardware resource-efficient to perform~\cite{simons2019review}. 
In this regard, the size of BNN models, even their original (i.e., not garbled) designs, is much smaller, which promotes further research into the implementation of NNs on hardware platforms, e.g., FPGAs.   
Having these advantages in mind, efforts have been made to improve the accuracy of BNNs, for instance, XNOR-Net has been proposed to narrow the gap when considering datasets like ImageNet~\cite{rastegari2016xnor}. 

By keeping the balance between the accuracy and hardware resource-efficiency in the original XNOR-Net design, it is also possible to achieve high hardware resources-efficiency, when garbling the BNNs as XNORing is free. 
To minimize the cost in terms of accuracy, one can take advantage of algorithmic solutions to scale up the size of XNOR-Net during the training phase and prune the network afterward cf.~\cite{riazi2019xonn}.  
Note that the accuracy of our garbled BNNs is not affected by the garbling process, i.e., the accuracy of the garbled NNs is similar to that of the trained NNs. 

\vspace{4pt}\noindent\textbf{Homomorphic Encryption (HE) vs. GCs.}
One of the promising solutions for oblivious inference is HE, where fully HE enables us to perform computation on encrypted data at the price of (relatively) high computational complexity. 
Partially HE (PHE) exhibits less overhead, but solely supports a subset of functions or depth-bounded arithmetic circuits. 
Even though HE-based DL accelerators could suffer from difficulties, e.g., complicated implementation of non-linear functions such as ReLU, high computational complexity, and truncation error~\cite{mohassel2017secureml,riazi2019xonn}, the advancement in this domain makes these accelerators more feasible in practice. 
In addition to software implementation of such accelerators, including~\cite{mohassel2017secureml,liu2017oblivious,gilad2016cryptonets,juvekar2018gazelle}, a hardware accelerator has been designed to bridge the gap between FHE and unencrypted computation through a specialized accelerator taking advantage of hardware acceleration~\cite{samardzic2021f1}.  

Since HE can also offer resiliency against SCA, it could be useful to compare the resource utilization of HWGN\textsuperscript{2} and a HE-based DL accelerator. 
As results for typical NNs employed in the relevant work, e.g., ~\cite{mohassel2017secureml,liu2017oblivious,juvekar2018gazelle}, are not reported in~\cite{samardzic2021f1}, we compare implementation of HWGN\textsuperscript{2} with ones presented in~\cite{liu2017oblivious,juvekar2018gazelle} (see Table~\ref{tab:GCvsHE}). 
The NN used in this example is the benchmark NN, first used in~\cite{mohassel2017secureml} and then employed against Gazelle and MiniONN in~\cite{liu2017oblivious,juvekar2018gazelle}.  
It is important to remark that these studies have focused on fully-software-based implementation of HE schemes, whereas HWGN\textsuperscript{2} results are collected from an ARTIX-7 FPGA (clock frequency of 20 MHz) board as the evaluator. 
As precisely mentioned in~\cite{juvekar2018gazelle}, both HE and GC techniques have their limitations. 
To select one for an application, it is suggested to take their trade-offs into account: for application with timing constraints, HWGN\textsuperscript{2} can offer similar time-complexity compared to Gazelle that shows a great deal of time-efficiency. 
On the other hand, if the major bottleneck is communication complexity, HWGN\textsuperscript{2} with the 1 instruction per OT could be useful. 
We remark again that the comparison presented in Table~\ref{tab:GCvsHE} is not completely fair as HWGN\textsuperscript{2} accelerator is implemented in hardware. 
Moreover, HE methods might not support circuit privacy as given by PFE and supported by HWGN\textsuperscript{2}, although there are HE protocols that attain this objective~\cite{bourse2016fhe,gentry2009fully}.  
As it is beyond the scope of this paper, we will not provide details on this. 
\begin{table}[t!]
\scriptsize
    \setlength{\tabcolsep}{0.4em}
    \renewcommand{\arraystretch}{0.1}
        \begin{center}
        \caption{Execution time and communication cost of HE-based approaches vs. HWGN\textsuperscript{2}. NNs (input layer:784 neurons, two hidden layers: 128 neurons each, and an output layer:10 neurons) trained on MNIST. The results for HE schemes presented in~\cite{liu2017oblivious,juvekar2018gazelle} are taken from~\cite{riazi2019xonn} (inst.: instructions).
}\label{tab:GCvsHE}
            \begin{tabular}{|c|c|c|}
                \hline
                Approach&Time (Sec)&Comminucation (MB)\\
                \hline
                \makecell{MiniONN*~\cite{liu2017oblivious}}&$1.04$&$15.8$
\\
                \hline
                \makecell{Gazelle*~\cite{juvekar2018gazelle}}&$0.09$&$0.5$
\\
                \hline
                \makecell{\textbf{HWGN\textsuperscript{2}} \textbf{(1 inst. per} \textbf{OT Interface)}}&$0.79$&$2.95$\\
                \hline
                \makecell{\textbf{HWGN\textsuperscript{2}} \textbf{(Complete set of inst. per} \textbf{OT interface)}}&\textbf{$0.14$}&\textbf{$148$}\\
                \hline
            \end{tabular}
    \end{center}
    \vspace{-5pt}
\end{table}

\vspace{4pt}\noindent\textbf{Interactive and non-interactive garbling: }
If an SFE protocol with garbling scheme $G$ is built upon an OT, after receiving the inputs $F$ and $d$, party~1 can initiate the OT protocol to obtain the output $y$. 
In other words, the interaction between parties happens solely when the OT protocol is executed. 
Therefore, if a non-interactive OT can be realized, the SFE protocol can be run non-interactively, e.g., by incorporating appropriate hardware such as trusted platform modules (TPMs)~\cite{gunupudi2008generalized}. 
For malicious security model, however, their proposal would not be secure; hence, One-Time Programs (OTP), i.e., GCs with OT calls being implemented with One-Time Memory (OTM) tokens, have been introduced~\cite{goldwasser2008one}. 
These primitives can also offer an important byproduct that is one-time or limited-time execution, i.e., the DL inference can occur only once or for a limited time, which can be helpful for enforcing license agreements.
To the best of our knowledge, J\"{a}rvinen et al.~\cite{jarvinen2010garbled} was the first to adopt these notions and further obtained leakage-resilience hardware implementations of cryptographic implementations, even in the non-interactive setting and in the presence of malicious adversary. 
Nevertheless, in their design, the privacy of the circuit has not been dealt with. 
The extension of HWGN\textsuperscript{2} to non-interactive model can be considered as our future work. 
\vspace{-10pt}

%% file: 7.Conclusion.tex
\section{Conclusion}\label{sec:conclusion}
In this paper, we have examined the feasibility of garbling to prevent attackers from launching SCA attacks against DL hardware accelerators.
We have implemented HWGN\textsuperscript{2} as a garbled DL hardware accelerator on an Artix-7 FPGA. 
By tailoring the concepts known only for software garbled DL accelerator~\cite{hussain2020tinygarble2} to the needs of a hardware DL accelerator,  the implementation of such accelerator is enhanced: HWGN\textsuperscript{2} requires up to $62.5\times$ fewer logical and $66\times$ less memory utilization compared to the state-of-art approaches. 
This is indeed possible at the price of more communication overhead. 
HWGN\textsuperscript{2} provides users the flexibility to protect their NN IP both in real-time applications and in applications where the hardware resources are limited by hardware resource utilization or communication cost. 
As our leakage evaluation results indicated, for both EM and power side-channles, the t-scores are below the threshold ($\pm4.5$), which shows the side-channel leakage resiliency of HWGN\textsuperscript{2} with trace population in the order of millions.
Another strength of HWGN\textsuperscript{2} is the DL model architecture thanks to the SFE/PFE protocol realized through MIPS instructions. 
Last but not least, we have discussed how HWGN\textsuperscript{2} can be implemented in the presence of a malicious adversary-- as a by-product. 

%% file: 8.appendix.tex
\vspace{20pt}
\appendix 
\noindent\textbf{\large{Appendix A. HWGN\textsuperscript{2} in the Presence of Malicious Adversaries}}
\vspace{5pt}
 \label{flow:Malicious}



\begin{figure*}[t!]
\centering \noindent
\includegraphics[width=0.95\textwidth]{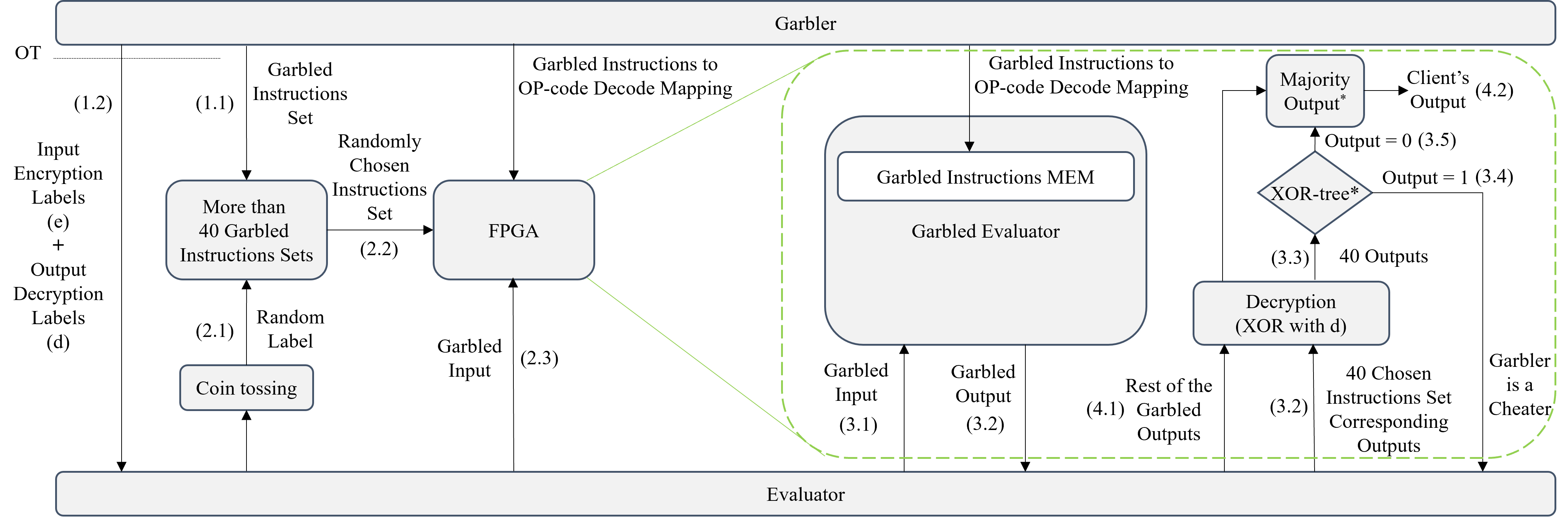}
\caption{The flow of the presented approach in the presence of a malicious adversary (Garbled Instructions Set: corresponding to the set of garbled MIPS instructions corresponding to the function $f$, Garbled Instructions to OP-code Decode Mapping: the complete garbled ISA based on which the garbled MIPS core evaluates the garbled MIPS instructions, Random Label: number of the randomly chosen garbled instructions set to be opened and checked, XOR-tree: a tree of cascaded XORs to find any difference in between the outputs corresponding to the evaluation of randomly chosen garbled instructions sets, and Majority Output: calculates the evaluator's output based on the majority of all evaluations outputs' value. (Note: starred modules (*) are inspired from~\cite{lindell2016fast}).}
\label{fig:Malicious_flow}
\end{figure*}

To implement our approach on FPGA, we have used garbled MIPS evaluator presented in Section~\ref{Garbled_MIPS}.
Figure \ref{fig:Malicious_flow} illustrates the flow of the HWGN\textsuperscript{2} in the presence of a malicious adversary.
Note that the numbers in the paragraphs are corresponding to the arrow indices in the Figure \ref{fig:Malicious_flow} and the garbled evaluator represents garbled MIPS evaluator.
In the first step (1.1), the garbler constructs many garbled instructions sets, more than 40 instruction sets, from function $f$ and sends them to the evaluator.
In addition, the garbler generates two matrices (1.2): $e$ and $d$ corresponding to constructed garbled instructions sets' input encryption labels output decryption labels, respectively.
The dimension of each matrix, $e$ and $d$, equals $s\times128$ where $s$ is the number of constructed garbled instructions sets.
In the next step (2.1), the evaluator generates 40 random labels based on the coin-tossing technique.
Based on these 40 random labels, 40 garbled instructions sets are chosen, opened, and checked by the evaluator (2.2).
It should be noted that to check all of these 40 garbled instructions sets, the evaluator must garbled and feed the same input to the evaluator (2.3), to check whether every 40 garbled instructions sets generate the same output or not.
The opening and checking process contains two phases: (i) evaluation and (ii) XOR-tree approach presented in~\cite{lindell2007efficient}. 
In the evaluation phase, the evaluator encrypts the chosen input with the corresponding input encryption labels ($e$) (3.1), evaluates the chosen garbled instructions set, and decrypts the generated garbled output (3.2) using the corresponding output decryption labels ($d$) (3.3).
Thereafter, in the XOR-tree phase, the evaluator calculates the XOR value of all decrypted outputs (3.3), bit by bit, to find any difference between the outputs. 
If the output of the XOR-tree becomes 1 (3.4), it means that the garbler is a cheater.
Otherwise (3.5), the evaluator can only assume that none of the 40 chosen garbled instructions sets is corrupted.
To minimize the cheating probability of the garbler even more, the evaluator evaluates the rest of the garbled instructions sets (4.1) (the remaining sets of instructions in (1.1)) and chooses the majority output of the evaluation of all the garbled instructions sets as its correct output (4.2).

The above-mentioned protection approach against the malicious garbler introduces two overheads to the design: (i) the XOR-tree and Majority Output calculation modules and (ii) increasing the size of garbled instructions memory of garbled evaluator core.
To alleviate the latter overhead, HWGN\textsuperscript{2} with an improved hardware resource efficiency is a better choice to minimize the hardware memory resource utilization with the trade-off of increasing communication costs.
In a nutshell, in real-time applications, the hardware resource utilization should be maximum and the communication cost should be minimum to obtain the minimum execution time; using TinyGarble-based implementation of HWGN\textsuperscript{2} is the best implementation choice.
On the contrary, in the application where the hardware resource utilization is the burden, the design netlist must be divided into $n$ sub-netlists, where $n$ is the number of gates which means each sub-netlist contains just one gate, and the evaluation of GC must be done gate by gate each via an OT interaction (refer to Section \ref{Implementation_Evaluation} to see the comparison results).
Overall, using HWGN\textsuperscript{2} gives the flexibility of using GCs in different types of applications. 

\noindent\textbf{Hardware Resource Utilization in the Presence of a Malicious Adversary.} 
The implementation flow of HWGN\textsuperscript{2} in the presence of a malicious adversary is similar to the implementation in the case of an  adversary with two extra modules: (i) XOR-tree and (ii) Majority Output.
Table \ref{Tab_5} shows HWGN\textsuperscript{2} in the presence of a malicious adversary overall and each module's hardware resource utilization.
The overhead of XOR-tree and Majority Output modules are negligible compared to the garbled MIPS evaluator.
The only noticeable overhead is the number of OT interactions.
As described in Section \ref{flow:Malicious} and mentioned in~\cite{lindell2016fast}, in the case of a malicious adversary, the garbler must construct more than $40$ GCs ($s$ in the last row of Table \ref{Tab_5}) which increases the OT interactions requirement of the securing HWGN\textsuperscript{2} against a malicious garbler up to $s$ times where $s$ is the number of GCs constructed by the garbler. HWGN\textsuperscript{2} in the presence of a malicious adversary memory footprint follows $O(s\times(I+N_{gate,m}+i_m))$ where $s$ is the number of GC constructed by garbler with $s$ is greater than $40$~\cite{lindell2016fast}. 

\begin{table}[t!]
\scriptsize
    \setlength{\tabcolsep}{0.4em}
    \renewcommand{\arraystretch}{0.1}
        \begin{center}
        \caption{Hardware resource utilization costs of HWGN\textsuperscript{2} in the presence of a malicious adversary. NNs (input layer:784 neurons, three hidden layers: 1024 neurons each, and an output layer:10 neurons) trained on MNIST.}\label{Tab_5}
            \begin{tabular}{|c|c|c|c|}
                \hline
                Module&LUT&FF&OT interaction\\
                \hline
                \makecell{Garbled MIPS Evaluator}&$1775$&$1278$&$2346$\\
                \hline
                XOR-tree&$2$&$0$&$0$\\
                \hline
                \makecell{Majority Output}&$10$&$0$&$0$\\
                \hline
                \textbf{Overall}&\textbf{$1787$}&\textbf{$1278$}&\textbf{$2346\times S$}\\
                \hline
            \end{tabular}
    \end{center}

\end{table}

\begin{table}[t!]
\scriptsize
    \setlength{\tabcolsep}{0.4em}
    \renewcommand{\arraystretch}{0.1}
        \begin{center}
        \caption{Execution time and communication cost comparison between HWGN\textsuperscript{2} and the state-of-the-art approach in the presence of a malicious adversary (Implementation of LeNet inteface).}\label{Tab_8}
            \begin{tabular}{|c|c|c|}
                \hline
                Approach&Time (Sec)&Comminucation (MB)\\
                \hline
                \makecell{Tinygarble2~\cite{hussain2020tinygarble2}}&291&72.75\\
                \hline
                \textbf{HWGN\textsuperscript{2} (DL hardware accelerator)}&\textbf{$39.12$}&\textbf{$93.02$}\\
                \hline
                \textbf{HWGN\textsuperscript{2} (XNOR based DL hardware accelerator)}&\textbf{$26.43$}&\textbf{$61.83$}\\
                \hline
            \end{tabular}
    \end{center}

\end{table}

The execution time and communication cost comparison in the presence of a malicious adversary is given in Table \ref{Tab_8}.
To have a fair comparison, we have implemented LeNet~\cite{lecun1998gradient} on FPGA using HWGN\textsuperscript{2}.
The reason behind the execution time and communication cost differences between HWGN\textsuperscript{2} and Tinygarble2 is that the evaluation process is taking place on the FPGA for the former.
The computation power of FPGA allows us to obtain better performance compared to the software implementation of Tinygarble2.
However, still a vast amount of execution time and communication cost corresponsds to the OT interaction. 

\vspace{10pt}
\noindent\textbf{\large{Appendix~B. TinyGarble-based Implementation of \\ HWGN\textsuperscript{2}}}\label{Tinygarble-based Implementation}

\begin{figure}[t!]
\centering \noindent
\includegraphics[width=\columnwidth]{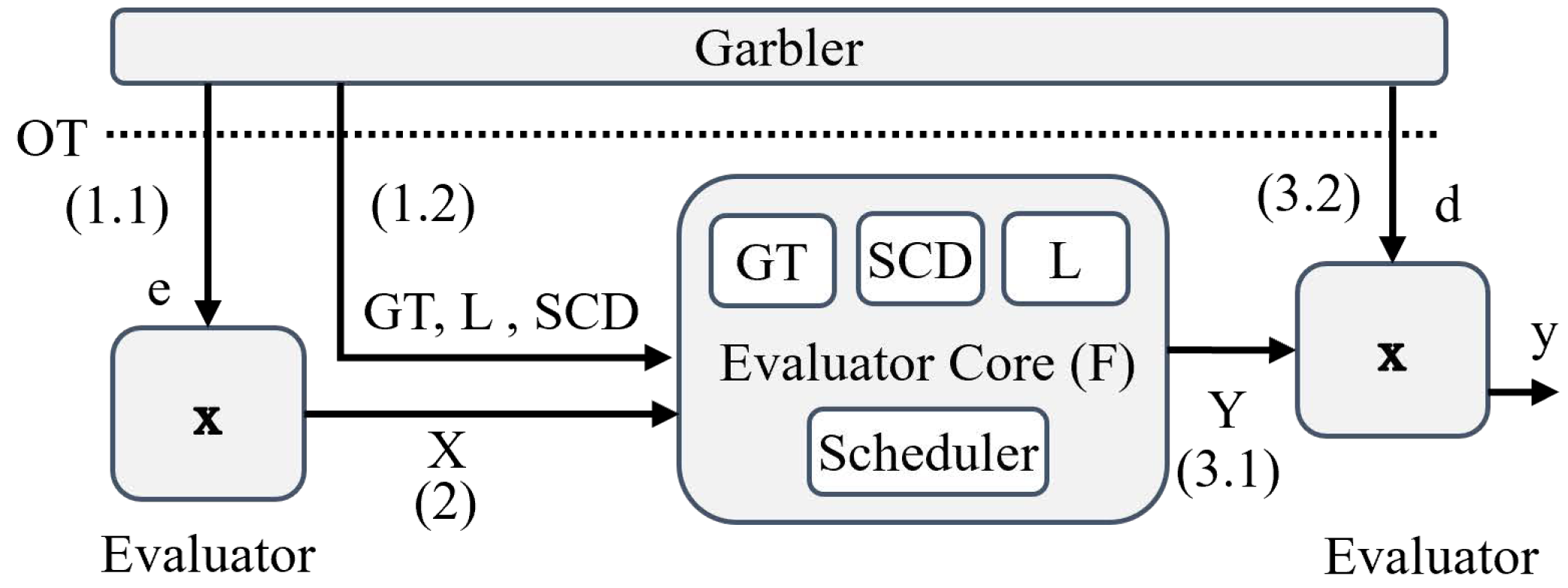}
\caption{TinyGarble-based implementation~\cite{songhori2015tinygarble} of HWGN\textsuperscript{2} ($L$: wires garbled labels, $GT$: garbled tables, $e$: encryption labels, $d$: decryption labels, $x$: evaluator's raw input, $X$: evaluator's garbled input, $Y$: garbled output, $Y_i$, $X_i$, $GT_i$, $L_i$: garbled input, output, garbled tables, wire labels corresponding to $i^{th}$ sub-netlist, respectively, $y$: evaluator's raw output, and $SCD$: A custom circuit description which allows TinyGarble to evaluate the Boolean circuit).}
\label{fig:HBC_flow}
\end{figure}

TinyGarble~\cite{songhori2015tinygarble} is a garbling framework that supports Yao's protocol and uses hardware-synthesis tools to generate circuits for secure computation automatically. 
The main advantage of TinyGarble is the scalability enabled by exploiting a sequential circuit description for garbled circuits and garbling optimization techniques such as Free-XOR~\cite{kolesnikov2008improved}, Row Reduction~\cite{naor1999privacy}, and Garbling with a Fixed-key Block Cipher~\cite{bellare2013efficient}. 
Figure~\ref{fig:HBC_flow}.a illustrates the flow of HWGN\textsuperscript{2} following TinyGarble~\cite{songhori2015tinygarble} approach.
At first, garbler chooses input encryption labels ($e$)  (Step~1.1) and constructs the GC of function $f$ by generating garbled tables ($GT$) of all gates, garbled labels ($L$) of all wires, and a custom circuit description ($SCD$) file (Step~1.2), which is the mapping between the GC and function $f$. 
$GT,L,SCD,e$ are sent through one OT interaction to the evaluator for further garbling protocol process. 
$e$ is then used by the evaluator to generate garbled input $X$ from the evaluator's input $x$ (Step~2). 
Afterward in Step~3.1, $GT$, $L$, and $SCD$ are used by the evaluator to evaluate the GC based on the given $X$ sequentially using the scheduler module (cf.~\cite{songhori2015tinygarble} for more information).
In the final step (Step~3.2), output decryption labels ($d$) are sent to the evaluator to decrypt the evaluator core's garbled output $Y$ and obtain its raw output $y$.
The sequential evaluation supported by TinyGarble provides the GC protocol the scalability of evaluation of larger netlists.
However, when one implements the DL hardware accelerator in the garbled format which has a large netlist, memory and logical resource utilization become burdens for DL hardware accelerators~\cite{jarvinen2010garbled} (see Section~\ref{Implementation_Evaluation}). 

\begin{figure}[t!]
\centering \noindent
\includegraphics[width=\columnwidth]{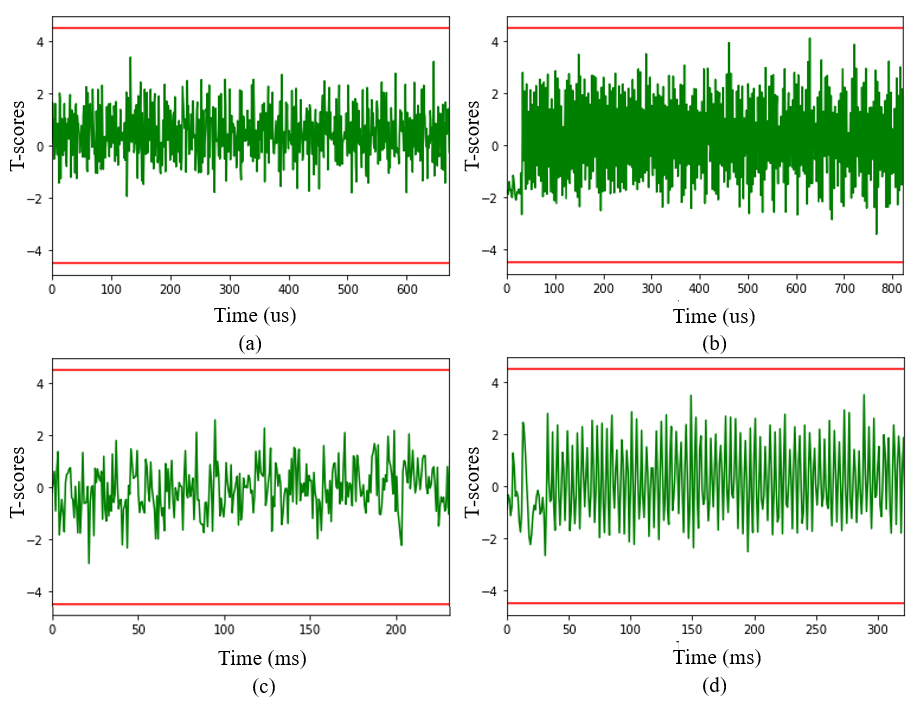}
\caption{TVLA test results (a) HWGN\textsuperscript{2} applied to XNOR-based BM2 (capacity whole set of instructions per OT) (b) BM2 with the capacity of complete set of instructions per OT interaction, (c) HWGN\textsuperscript{2} applied to XNOR-based BM2 (capacity 1 instruction per OT), and (d) BM2 with the capacity of 1 instruction per OT (calculated for $2M$ EM traces).}
\label{fig:garbled_capacity_comparison_2M_EM}
\end{figure}

\vspace{10pt}
\noindent\textbf{\large{Appendix C. TVLA Test Evaluation of EM Side-channel}}\label{TVLA_EM}
One of the first studies that has compared the capabilities of attackers launching power vs. EM SCA is~\cite{peeters2007power}, where it is suggested that the EM leakage can provide more information than the power consumption of the same chip cf.~\cite{standaert2008using}. 
This has been further justified in~\cite{standaert2008using} through the evaluation of the information theoretic and security metrics~\cite{standaert2009unified}. 
Therefore, it might be thought that the EM side-channel could offer some information about the secret, i.e., the weights of the garbled NN. 
To collect the EM traces, it has been already verified that measurements from the frontside of a chip can offer a high signal-to-noise ratio~\cite{heyszl2012strengths}; hence, we stick to this setting to perform measurements. 
Our setup described in Section~\ref{SC_measurement_setup} is equipped with HP EM probe 125 (SN126 0.2mm).
Figure~\ref{fig:garbled_capacity_comparison_2M_EM} shows the t-scores computed for HWGN\textsuperscript{2} applied against BM2. 
As shown in Figure~\ref{fig:garbled_capacity_comparison_2M_EM}, the t-scores of EM traces are below the threshold ($\pm4.5$) which is the proof of the EM leakage resiliency of HWGN\textsuperscript{2}.